\documentclass[preprint,12pt,sort&compress]{elsarticle}
\usepackage[utf8]{inputenc}
\usepackage{color}
\usepackage{graphicx}
\usepackage{placeins}
\usepackage[margin=1in]{geometry}
\usepackage{appendix}
\usepackage{amsmath, amssymb, bm}
\usepackage{multirow}
\usepackage{placeins}
\usepackage{xfrac}
\usepackage[colorlinks=true, linkcolor=blue, citecolor=blue]{hyperref}
\usepackage[nameinlink, capitalize]{cleveref}
\usepackage[font=footnotesize, labelfont=bf]{caption}
\usepackage{subcaption}
\usepackage[version=4]{mhchem}
\usepackage{lineno}
\usepackage{bm}
\usepackage{comment}
\usepackage[T1]{fontenc}
\usepackage{anyfontsize}

\usepackage{tablefootnote}
\usepackage{float}
\usepackage{stackengine}
\usepackage{mathrsfs} 
\usepackage{graphicx} 
\usepackage{booktabs}

\usepackage{siunitx}
\journal{Journal of Nuclear Materials}

\begin{document}

\begin{frontmatter}

\title{Modeling of silver transport in cubic SiC: Integrating molecular dynamics, bounds averaging, and uncertainty quantification}

\author[ncsu]{Mohamed AbdulHameed}
\author[ncsu,ent]{Khadija Mahbuba}
\author[psu]{Mahmoud Yaseen}
\author[ncsu]{Amr Ibrahim}
\author[epri]{Daniel Moneghan}
\author[ncsu,inl]{Benjamin Beeler\corref{ca}}
\cortext[ca]{Corresponding author}
\ead{bwbeeler@ncsu.edu}

\address[ncsu]{Department of Nuclear Engineering, North Carolina State University, Raleigh, NC 27695}
\address[inl]{Idaho National Laboratory, Idaho Falls, ID 83415}
\address[psu]{Department of Nuclear Engineering, Pennsylvania State University, University Park, PA 16802}
\address[ent]{Entergy Operations Inc.: Department of Fuels, St. Francisville, LA 70791}
\address[epri]{Electric Power Research Institute, 1300 West WT Harris Boulevard, Charlotte, NC 28262}

\begin{abstract}

Silver released from TRISO fuel particles can migrate through the SiC layer and deposit on reactor components, posing radiation hazards and operational challenges. Despite numerous proposed mechanisms, the precise pathway of silver transport through intact 3C-SiC remains unresolved. We present a physics-informed model for estimating the effective diffusivity of silver in polycrystalline 3C–SiC. Molecular dynamics (MD) simulations yield diffusivities for $\Sigma 3$ and $\Sigma 9$ grain boundaries (GBs), while literature values are used for other GB types and the bulk. These are combined using a bounds-averaging approach accounting for distinct GB transport properties. Bayesian inference of experimental data provides credible intervals for effective Arrhenius parameters and reveals a correlation between activation energy and pre-exponential factor. Although the homogenized model captures GB-mediated transport mechanisms, it overpredicts silver diffusivity relative to experiments. To resolve this, a multiplicative correction based on reversible trapping at nano-pores is introduced. It is derived from first principles and is shown to reproduce observed transport behavior. Sensitivity analysis identified trap desorption energy and $\Sigma 9$ GB diffusivity as dominant factors influencing Ag transport. The resulting framework provides a mechanistic description of Ag transport suitable for integration into higher-scale fuel performance models.

\end{abstract}

\begin{keyword}
Silver transport \sep Silicon carbide \sep Molecular dynamics \sep Sensitivity analysis \sep Uncertainty quantification 
\end{keyword}

\end{frontmatter}

\newpage


\section{Introduction}


The tri-structural isotropic (TRISO) coated fuel particles represent an advanced class of nuclear fuel engineered for advanced reactors. They consist of a microspherical fuel kernel coated with successive layers of porous pyrolytic carbon (PyC), inner dense PyC (IPyC), silicon carbide (SiC), and outer dense PyC (OPyC) \cite{Minato2020}. Among these layers, SiC serves a dual role: It (\textit{a}) provides structural support against irradiation-induced stresses in the PyC layers, and (\textit{b}) acts as a barrier to the diffusion of fission and activation products, e.g., palladium (Pd) and silver ($^{\text{110m}}$Ag), which easily penetrate the IPyC layer \cite{Snead2007, Verfondern2020}. Note that $^{109}$Ag is a fission product from which the activation product $^{\text{110m}}$Ag, a strong $\gamma$-ray emitter with a half-life of about 250 days, is produced via the $^{109}$Ag $(n, \gamma)$ $^\text{110m}$Ag reaction \cite{Lopez-Honorato2010, VanRooyen2014b}.

The SiC polymorph employed in the TRISO fuel particles is $\beta$-SiC, more commonly known as 3C-SiC, where ``3'' refers to the three-bilayer (ABCABC...) stacking viewed along the $\langle 111 \rangle$ direction, and ``C'' indicates the cubic symmetry of the crystal \cite{Shaffer1969}. 3C-SiC is the only cubic polymorph of SiC. It exhibits a zinc-blende crystal structure and a room temperature lattice constant of 4.36 {\AA} \cite{Sultan2022}. High-purity, stoichiometric 3C-SiC for TRISO applications is typically synthesized using chemical vapor deposition (CVD) techniques \cite{Snead2007}.

Experimental investigations have revealed that fission and activation products, especially Pd and Ag, can migrate from the fuel kernel to the inner surface of the SiC layer, potentially compromising its structural integrity \cite{Verfondern2020}. Of particular concern is silver, which has been found to migrate through seemingly intact TRISO particles and subsequently be released into the primary circuit \cite{Verfondern2020,VanRooyen2014b}. Such silver release results in its deposition on high-maintenance components, thereby increasing radiation exposure risks for maintenance personnel \cite{VanDerMerwe2009}. Addressing the associated operational challenges may necessitate either costly maintenance strategies or restrictive operational constraints, such as limiting reactor power output and outlet temperature \cite{VanDerMerwe2009}. However, such temperature reductions contradict the primary objective of high-temperature reactors, which is to operate at elevated temperatures for enhanced thermal efficiency.

Despite decades of research, no consensus has been reached regarding the mechanism of Ag transport through intact 3C-SiC, with various proposed mechanisms lacking definitive experimental or computational validation \cite{VanRooyen2014}. These mechanisms include grain boundary (GB) diffusion, surface diffusion, vapor transport through interconnected nano-pores or nano-cracks, Pd-assisted GB transport, and irradiation-enhanced bulk diffusion \cite{VanRooyen2014, VanRooyen2014b}. Among these, GB diffusion is widely regarded as the most likely plausible pathway for Ag transport \cite{VanRooyen2014}. Nevertheless, a comprehensive investigation elucidating the atomistic details of this mechanism remains absent.

Olivier, Neethling, and O'Connell \cite{Olivier2012, Neethling2012, Olivier2013, OConnell2014} conducted several experiments involving polycrystalline 3C-SiC in contact with Pd and Ag. They proposed a mechanism whereby Pd preferentially penetrates GBs in polycrystalline 3C-SiC, inducing corrosion and the formation of palladium silicides. Silver then reacts with these silicides to form Ag-rich Pd-Ag-Si solid solutions that migrate along GBs as nodular structures. These nodules advance by dissolving 3C-SiC at their leading edge and reprecipitating it at the trailing edge. Notably, only minimal Pd concentrations are required to facilitate this transport, which is further accelerated under high-temperature neutron irradiation \cite{OConnell2014}. Because pure Ag was not observed to penetrate 3C-SiC, they suggested that Ag diffusion through GBs requires the presence of Pd \cite{Neethling2012}. However, a critical limitation of these studies lies in their use of Pd and Ag concentrations significantly higher than those present in TRISO particles, possibly leading to different diffusion mechanisms \cite{VanRooyen2014b}.

In contrast, López-Honorato \textit{et al.} \cite{Lopez-Honorato2010, Lopez-Honorato2011} observed Ag transport in the absence of Pd, thereby contradicting Olivier and Neethling's findings. They proposed that Ag migrates through a sequence of GBs and nano-voids, as nano-voids introduced during CVD are not interconnected and are separated by at least one grain \cite{Lopez-Honorato2010}. Their work also emphasized that the relatively straight GBs found in columnar grains facilitate fission product diffusion, whereas the more convoluted boundaries in fine-grained structures significantly hinder it \cite{Petti2003, Lopez-Honorato2010}.

To reconcile these conflicting observations, Olivier and Neethling \cite{Olivier2013} speculated that Ag vapor transport might have occurred in López-Honorato \textit{et al.}’s experiments, which were conducted at 1573~K---above Ag's melting point. López-Honorato \textit{et al.} further demonstrated that the presence of nano- and micro-pores at GBs can enhance Ag diffusion coefficients by at least two orders of magnitude \cite{Lopez-Honorato2011}. While porosity can promote Ag diffusion or even shift the transport mechanism to vapor-phase transport, its presence alone does not necessarily result in significant diffusion \cite{Lopez-Honorato2011, VanRooyen2014}. That is, defects within the grain structure can contribute to Ag migration via defect-GB-defect pathways; however, due to their separation by hundreds of nanometers, GBs typically dominate the diffusion process \cite{Lopez-Honorato2011}.

Ideally, a predictive understanding of Ag transport in 3C-SiC requires knowledge of the GB types present in irradiated SiC and the corresponding silver diffusivities. However, most computational studies to date have either focused on defect energetics and Ag migration in bulk 3C-SiC \cite{Shrader2011, Jiang2021} or examined a limited set of GBs, often selected arbitrarily or based on crystallographic simplicity, without regard to their actual abundance in irradiated microstructures \cite{Khalil2011, Aagesen2022, Ko2016}. Notably, $\Sigma 3$ and $\Sigma 9$ coincidence-site lattice (CSL) boundaries, known for their prevalence in SiC \cite{Lillo2016,Kirchhofer2013,Randle2011}, have not been systematically incorporated into Ag transport models. Lillo and van Rooyen \cite{Lillo2016} characterized GBs in irradiated SiC using precession electron diffraction and found that $\Sigma 3$ GBs comprise nearly half of all GBs, while $\Sigma 9$ and $\Sigma 5$ boundaries, though less abundant, still occur in non-negligible fractions. Moreover, chemical vapor deposition (CVD), the method typically used to fabricate TRISO SiC coatings, promotes the formation of CSL boundaries due to growth-induced texture \cite{Tanaka2002}. This disconnect between modeling assumptions and experimental microstructures limits the development of microstructure-informed models of Ag transport in SiC.

In this work, we present a physically grounded framework for estimating the effective diffusivity of Ag in cubic SiC by combining experimentally measured GB-type distributions with silver diffusivity data across different GB types. Literature values are used for bulk diffusion, high-angle GBs (HAGBs), and $\Sigma 5$ boundaries (as detailed in \cref{Sec:Current}), while molecular dynamics simulations are employed to calculate Ag diffusivities in $\Sigma 3$ and $\Sigma 9$ GBs. These data are integrated in a bounds-averaging scheme to obtain a closed‑form expression for the effective diffusivity, $D_\text{eff}(T)$. A subsequent Bayesian analysis is used to infer the most probable effective diffusion parameters, $Q_\text{eff}$ and $D_{0,\text{eff}}$, based on available experimental measurements. Finally, a local derivative-based sensitivity analysis identifies the Arrhenius parameters that most strongly influence $D_\text{eff}(T)$. Together, this workflow offers a systematic, microstructure-informed approach to modeling Ag transport in 3C-SiC.


\section{Methodology}

\subsection{Effective diffusivity model}

To estimate the effective diffusivity of Ag in 3C-SiC, we adopt a method based on averaging the theoretical upper and lower bounds of diffusivity. In our approach, the microstructure of the material is idealized as a collection of distinct, homogeneous phases, with the bulk and each GB type treated as separate phases. Chen~\cite{ChenThesis} reported analytical expressions for the upper and lower bounds of effective diffusivity in a two-phase system. We extend this formalism to systems comprising multiple phases, which is essential for materials such as 3C-SiC, where multiple GB types with different diffusivities coexist. These bounds are derived by analogy to the well-established rules for combining electrical conductances or thermal conductivities \cite{Dong2015} in parallel and series configurations.

The upper bound of the effective diffusivity, $D_u$, corresponds to the case where all phases contribute independently and simultaneously to the total diffusion flux, akin to conductors in parallel. This scenario assumes no resistance to diffusion across phase boundaries. Here, the effective diffusivity is given by the weighted arithmetic mean:
\begin{equation}
D_u = \sum_i f_i D_i,
\label{Eq:DU}
\end{equation}
where $f_i$ denotes the volume fraction of phase $i$, and $D_i$ is its intrinsic diffusivity.

In contrast, the lower bound of the effective diffusivity, $D_l$, represents a configuration where the diffusion paths traverse the different phases sequentially, similar to conductors arranged in series. This yields a weighted harmonic mean of the diffusivities:
\begin{equation}
D_l = \left( \sum_i \frac{f_i}{D_i} \right)^{-1}.
\label{Eq:DL}
\end{equation}

These two bounds bracket the range within which the true effective diffusivity of the composite system is expected to lie. Inspired by the Voigt-Reuss-Hill average of elastic moduli \cite{AbdulHameed2024}, we take the average of $D_u$ and $D_l$ to obtain a practical estimate for the effective diffusivity of Ag in 3C-SiC.
\begin{equation}
D_\text{eff} = \frac{D_u + D_l}{2} = D_\text{0,eff} \exp\left(-\frac{Q_\text{eff}}{kT}\right).
\label{Eq:Dav}
\end{equation}

To assess the validity of our approach, we compare it against the percolation theory-based model proposed by Chen and Schuh~\cite{Chen2007}. Their model considers two phases, bulk and a single GB type, and yields the following expression for the effective diffusivity:
\begin{equation}
D_\text{eff} = f D_\text{GB} + (1-f) D_b + \frac{ f (1-f) ( D_\text{GB} - D_b )^2 }{ f ( D_\text{GB} - D_b ) - 3 D_\text{GB} }.
\label{Eq:Chen}
\end{equation}
In this equation, $f$ is the volume fraction of GBs, $D_\text{GB}$ is the GB diffusivity, and $D_b$ is the bulk diffusivity. The first two terms correspond to the upper bound expression for a two-phase system as given in \cref{Eq:DU}, while the third term introduces a correction that accounts for the deviation from the purely parallel approximation. This correction term is always negative, thereby reducing the overall diffusivity compared to the upper bound.

To enable a meaningful comparison, we apply both methods, our bounds-averaging method (\cref{Eq:Dav} denoted Method~1) and the percolation-based model of Chen and Schuh (\cref{Eq:Chen} denoted Method~2), to a two-phase system consisting of bulk material and a single isotropic GB type. For this configuration, the expressions for the upper and lower bounds in our method simplify to:
\begin{equation}
D_u = f_\text{GB} D_\text{GB} + (1-f_\text{GB}) D_b, \quad D_l = \left( \frac{f_\text{GB}}{D_\text{GB}} + \frac{1-f_\text{GB}}{D_b} \right)^{-1}.
\end{equation}
As before, we take the arithmetic mean of $D_u$ and $D_l$ to estimate the effective diffusivity.

The GB volume fraction $f_\text{GB}$ is computed as \cite{Chen2007}:
\begin{equation}
f_\text{GB} = \frac{H \delta}{d},
\end{equation}
where $\delta$ is the GB width (assumed to be 1~nm), $d$ is the average grain size (taken to be 1~$\mu$m, as typical for SiC in TRISO fuel particles~\cite{Lopez-Honorato2010,Rohbeck2016}), and $H$ is a geometric shape factor. For mono-sized truncated octahedral grains, $H = 2.6575$, while for Voronoi polyhedral grains, $H = 2.9105$~\cite{Chen2007}. Assuming the latter, we obtain $f_\text{GB}$ = 0.0029.

\begin{figure}[h!]
 \centering
 \includegraphics[width=0.8\textwidth]{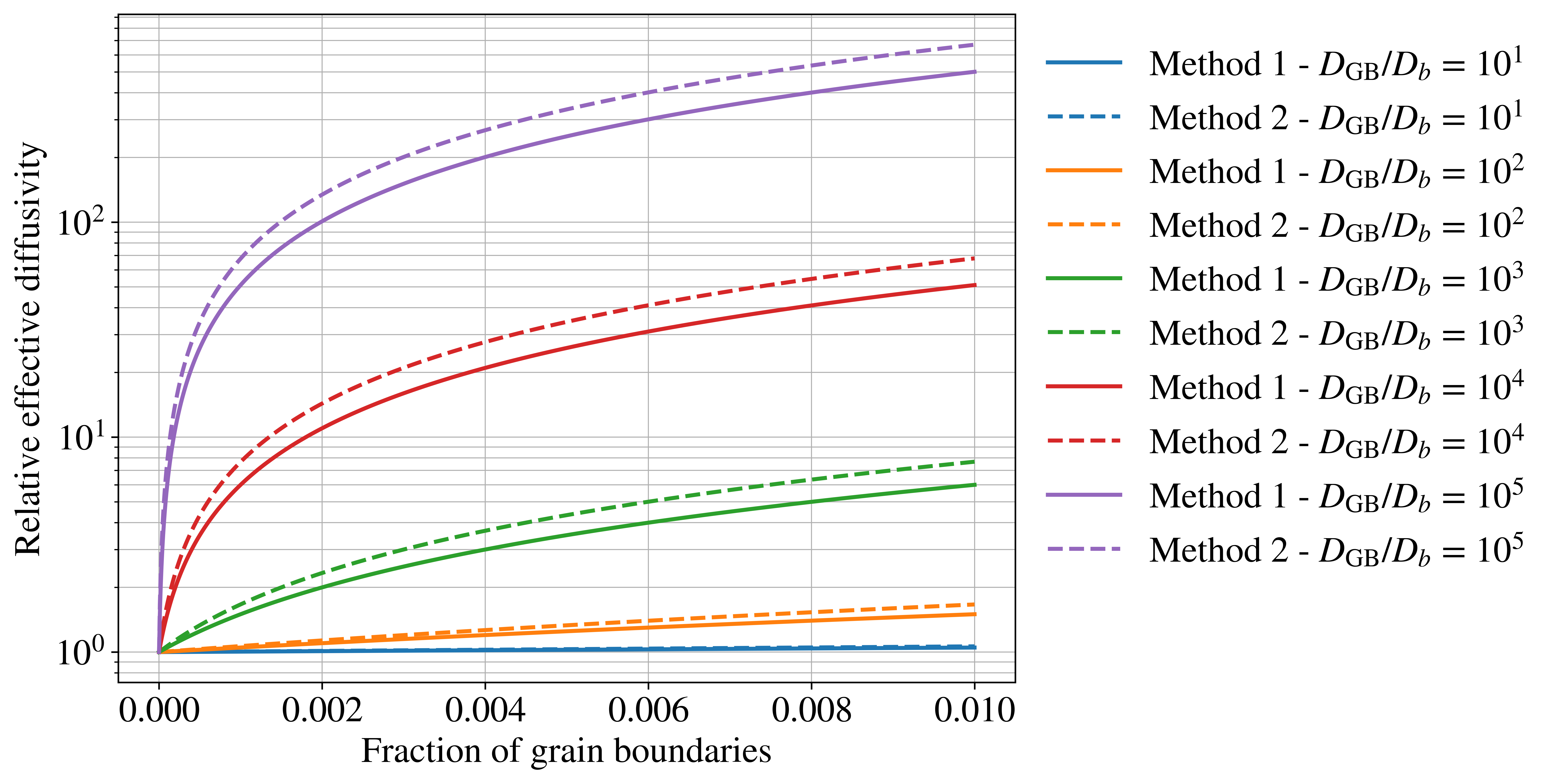}
 \caption{(Color online) Relative effective diffusivity computed using our averaging method (Method 1) and the percolation-based method by Chen and Schuh~\cite{Chen2007} (Method 2), as a function of the GB-to-bulk diffusivity ratio $D_\text{GB}/D_b$.}
 \label{Fig:DGB_Db}
\end{figure}

We evaluate both models over a wide range of diffusivity ratios, $D_\text{GB}/D_b$ = 10--$10^5$, with the bulk diffusivity fixed at $D_b = 1$. \cref{Fig:DGB_Db} presents the results of this comparison. Across the entire range of $D_\text{GB}/D_b$ values considered, our method yields slightly lower estimates than the percolation-based model. However, the difference remains within a factor of 2 for all values of $f$, which is negligible on a logarithmic scale. This strong agreement supports the validity of our method in the isotropic two-phase case.

The principal advantage of our approach lies in its extensibility. Unlike the Chen and Schuh model, which is restricted to a single isotropic GB type, our method can be naturally generalized to systems with multiple anisotropic GB types, each characterized by a distinct diffusivity. We therefore argue that the validation of our method in the isotropic limit provides a foundation for its application in more complex, anisotropic microstructures.


\subsection{Existing data on GB types and Ag diffusion in SiC}

To characterize GBs in SiC belonging to irradiated TRISO particles, Lillo and van Rooyen \cite{Lillo2016} employed precession electron diffraction in a transmission electron microscope. They reported the following distribution of GB types: random high-angle GBs (HAGBs) (\cref{Tab:GB}): 30\%, $\Sigma 3$ GBs: 46\%, $\Sigma 9$ GBs: 5\%, $\Sigma 27$ GBs: 1\%, other coincidence-site lattice (CSL) GBs: 7\%, and low-angle GBs: 11\%. As expected, $\Sigma 9$ GBs are the second most abundant boundary type after $\Sigma 3$ because they are geometrically necessary byproducts of $\Sigma 3 + \Sigma 3$ interactions at triple junctions \cite{Randle2011}. It is noted that $\Sigma 5$ GBs, belonging to the ``other CSL GBs'' category, seemingly exist in a percentage comparable to (or slightly lower than) that of $\Sigma 27$ GBs. In this context, reported CSL GBs are those with $\Sigma \leq 29$. The high fraction of CSL GBs is expected since specimens obtained via CVD often exhibit a preferred orientation during growth, which increases the probability of CSL GB formation \cite{Tanaka2002}. In this work, we rely on measurements by Lillo and van Rooyen \cite{Lillo2016} in irradiated 3C-SiC. GB volume fractions in 3C-SiC have also been measured by Kirchhofer \textit{et al.} \cite{Kirchhofer2013}, as shown in \cref{Tab:GB}. Except for the variation of $\Sigma 3$ and HAGBs, the values do not differ much. Moreover, the study by Kirchhofer \textit{et al.} is done on unirradiated 3C-SiC, and the data by Lillo and van Ryoon are more relevant to the problem under study.

\begin{table}[h!]
\centering
\footnotesize
\caption{Percentages of different GB types in irradiated 3C-SiC as characterized by Lillo and van Rooyen \cite{Lillo2016}, and unirradiated 3C-SiC as characterized by Kirchhofer \textit{et al.} \cite{Kirchhofer2013}. Values preceded by ``$\sim$'' are estimated by simple subtraction and are not given explicitly in the relevant studies. In this context, CSL GBs are those with $\Sigma \leq 29$.}
\begin{tabular}{lcc}
\hline
GB type & Irradiated \cite{Lillo2016} & Unirradiated \cite{Kirchhofer2013} \\
\hline
Random high-angle GBs & 30\% & 39\% \\
$\Sigma 3$ GBs & 46\% & 34.5\% \\
$\Sigma 9$ GBs & 5\% & 5.25\% \\
$\Sigma 27$ GBs & 1\% & 1.25\% \\
$\Sigma 5$ + Other CSL GBs & $\sim 1\%$ + $\sim 6\%$ & $\sim 7.25\%$ \\
Low-angle GBs & 11\% & 12.75\% \\
\hline
\end{tabular}
\label{Tab:GB}
\end{table}

We do not have diffusivity data for all GB types listed in \cref{Tab:GB}. Therefore, we employ an approximation: The $\Sigma 9$ and $\Sigma 27$ GBs are grouped into a single type, denoted as $\Sigma 9 / \Sigma 27$, under the assumption that they share the same diffusivity. Their corresponding volume fractions are thus combined. A similar approach is applied to $\Sigma 5$ and the remaining CSL GBs, which are collectively treated as a single GB type. Additionally, LAGBs are assumed to exhibit bulk diffusivity. These simplifications are expected to have a small impact on the effective diffusivity $D_\text{eff}$, owing to the small volume fractions of the merged GB types. This effectively leads to 6\% of GBs with the diffusivity of $\Sigma 9$, 7\% with the diffusivity of $\Sigma 5$, and 11\% with the diffusivity of the bulk, while the volume fractions of $\Sigma 3$ and HAGBs are kept at 46\% and 30\%, respectively (see \cref{Tab:Values}).

Computational studies of Ag diffusion in 3C-SiC have primarily focused on bulk or select GBs. Kohler \cite{Kohler2002} and Wojdyr \textit{et al.} \cite{Wojdyr2010} used molecular dynamics (MD) and Tersoff potential parametrizations to investigate tilt GBs and antiphase boundaries. Shrader \textit{et al.} \cite{Shrader2011} applied density functional theory (DFT) to study Ag defect energetics and migration pathways in bulk 3C-SiC. Jiang \textit{et al.} \cite{Jiang2021} conducted a more comprehensive DFT and kinetic Monte Carlo (KMC) study of Ag diffusion in bulk 3C-SiC over the temperature range 1073--2073~K, yielding the following expression for the diffusivity of Ag$_\text{C}$:
\begin{equation}
D_b = 2.4 \times 10^{-4} \exp\left(- \frac{5.34 \pm 0.58 \ \text{eV}}{k T} \right).
\end{equation}

Jiang \textit{et al.} also conducted MD calculations with 2-D and 3-D simulation boxes to estimate Ag diffusivity in HAGBs in 3C-SiC in the temperature range 2000--3000~K. They found that:
\begin{equation}
D_\text{HAGB} = 1.813 \times 10^{-7} \exp \left( - \frac{2.178 \text{ eV}}{k T} \right).
\end{equation}
It should be mentioned that Ko \textit{et al.} \cite{Ko2016} used an \textit{ab initio}-based KMC model to study Ag diffusion in amorphous SiC as a proxy for HAGBs, predicting:
\begin{equation}
D_\text{HAGB} = (2.73 \pm 1.09) \times 10^{-10} \exp\left(- \frac{2.79 \pm 0.18 \ \text{eV}}{k T} \right),
\end{equation}
which is 2-3 orders of magnitude lower than the reported Ag diffusivities extracted from integral release measurements. The approach adopted by Jiang \textit{et al.} is arguably more realistic, given that it models actual HAGBs rather than utilizing an amorphous structure as a surrogate.

Khalil \textit{et al.} \cite{Khalil2011} employed DFT to investigate the energetics and diffusion pathways of Ag point defects and Ag-containing defect clusters in the $\Sigma 3$ GBs of 3C-SiC. While their qualitative analysis provides valuable insight, a computational error in determining activation energies, specifically, the addition rather than subtraction of the binding energy, limits the quantitative reliability and interpretability of their results.

Aagesen \textit{et al.} \cite{Aagesen2022} employed MD simulations with an analytical bond-order potential to evaluate Ag diffusivity along the $\Sigma 5 \{210\}\langle 001\rangle$ GB in 3C-SiC between 1750--2750~K. Denoting the $\Sigma 5$ GB plane as the $xy$-plane, they found that for the $x$- and $y$-directions, the Ag diffusivities are:
\begin{align}
D_x &= 4.203 \times 10^{-10} \exp\left(-\frac{0.593 \ \text{eV}}{k T} \right), \\
D_y &= 6.161 \times 10^{-9} \exp\left(-\frac{0.954 \ \text{eV}}{k T} \right).
\label{Eq:DxDy}
\end{align}
As outlined in \ref{app}, we use \cref{Eq:DxDy} to derive the effective 2-D diffusivity of Ag in $\Sigma 5$ GBs as:
\begin{equation}
D_{\Sigma 5} = \frac{D_x + D_y}{2} = 2.567 \times 10^{-9} \exp\left(-\frac{0.847 \ \text{eV}}{k T} \right).
\label{Eq:Sigma5}
\end{equation}
Note that the findings of Aagesen \textit{et al.} suggest that Ag diffusion in well-ordered $\Sigma 5$ GBs is significantly faster than in random HAGBs, supporting earlier conclusions by López-Honorato \textit{et al.} regarding the impact of GB structure on diffusivity.


\subsection{Ag diffusion in $\Sigma 3$ and $\Sigma 9$ GBs}

\subsubsection{Atomic models of $\Sigma 3$ and $\Sigma 9$ GBs}

In the zinc-blende structure, GBs can be classified as \textit{polar} or \textit{non-polar} based on the stoichiometry in the interfacial region~\cite{Kohyama1991}. The distinction lies in the presence and type of \textit{wrong bonds}, which are homonuclear bonds such as Si-Si or C-C that are atypical in the ideal bulk structure~\cite{Kohyama1991}. In polar interfaces, only one type of wrong bond, or an excess of one kind, is found, whereas non-polar interfaces either contain no wrong bonds or an equal number of both kinds~\cite{Kohyama1994}.

This classification is particularly relevant when analyzing specific CSL boundaries. For example, for the coherent $\Sigma 3 \{ 111 \}$ and $\Sigma 9 \{ 122 \}$ GBs, it is possible to construct two polar interfaces and one non-polar interface, whereas for the incoherent $\Sigma 3 \{ 112 \}$ GB, only non-polar interfaces are feasible~\cite{Kohyama1994}. Note that a \textit{coherent} twin boundary, e.g., $\Sigma 3 \{ 111 \}$, is characterized by an interface that coincides with the twinning plane, whereas an \textit{incoherent} twin boundary, e.g., $\Sigma 9 \{ 112 \}$, does not lie on the twinning plane and is typically connected to a coherent twin boundary or to the boundaries of the twinned grain \cite{Novikov2003}. For example, in 3C-SiC, the $\Sigma 9 \{ 112 \}$ GB is usually perpendicularly connected to the $\Sigma 3 \{ 111 \}$ GB~\cite{Kohyama1991}. The structural characteristics of these boundaries vary significantly. The $\Sigma 3 \{ 111 \}$ GBs contain only 6-atom rings \cite{Kohler2002}, while reconstructed models of $\Sigma 3 \{112\}$ GBs contain 5-, 6- and 7-atom rings \cite{Kohyama1994,Tanaka2002}. $\Sigma 9 \{ 122 \}$ GBs contain 5- and 7-atom rings \cite{Kohyama1991}.

\begin{figure}[h!]
\centering
\begin{subfigure}{0.3\textwidth}
 \includegraphics[width=\textwidth]{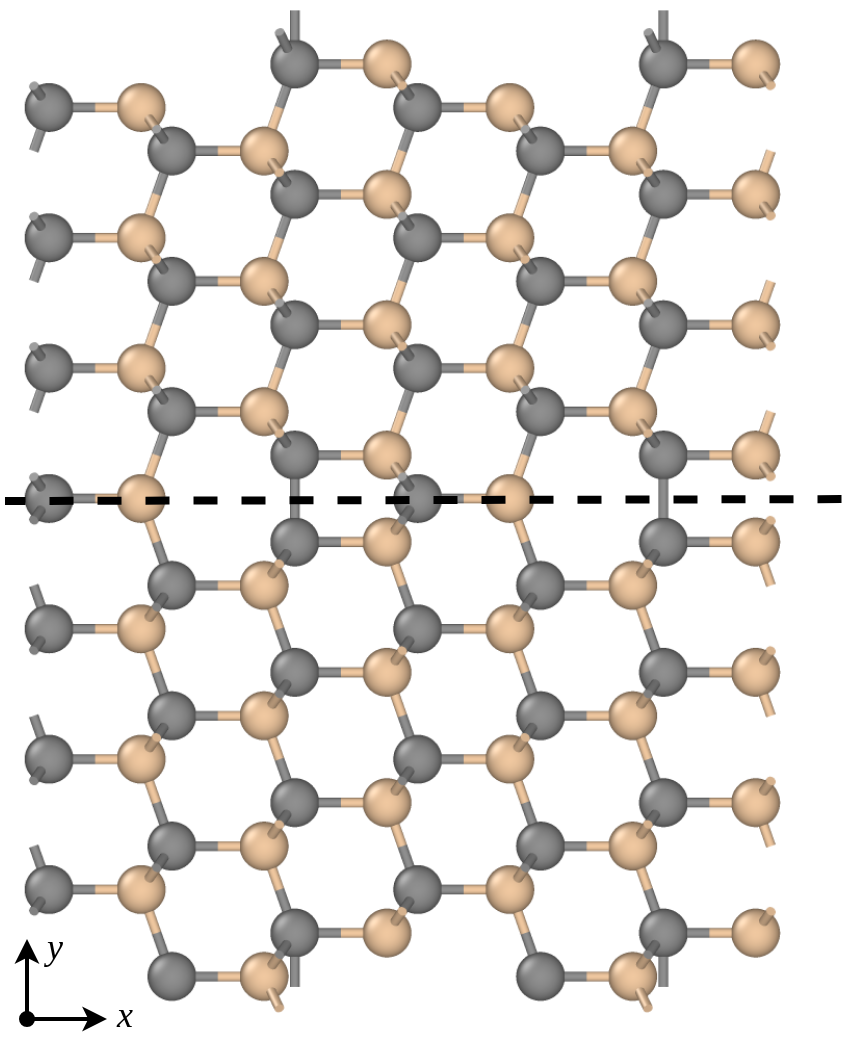}
 \caption{$p2'mm' (1 \times 1)$ model}
 \label{Fig:S3-112-1}
\end{subfigure}
\hspace{2em}
\begin{subfigure}{0.3\textwidth}
 \includegraphics[width=\textwidth]{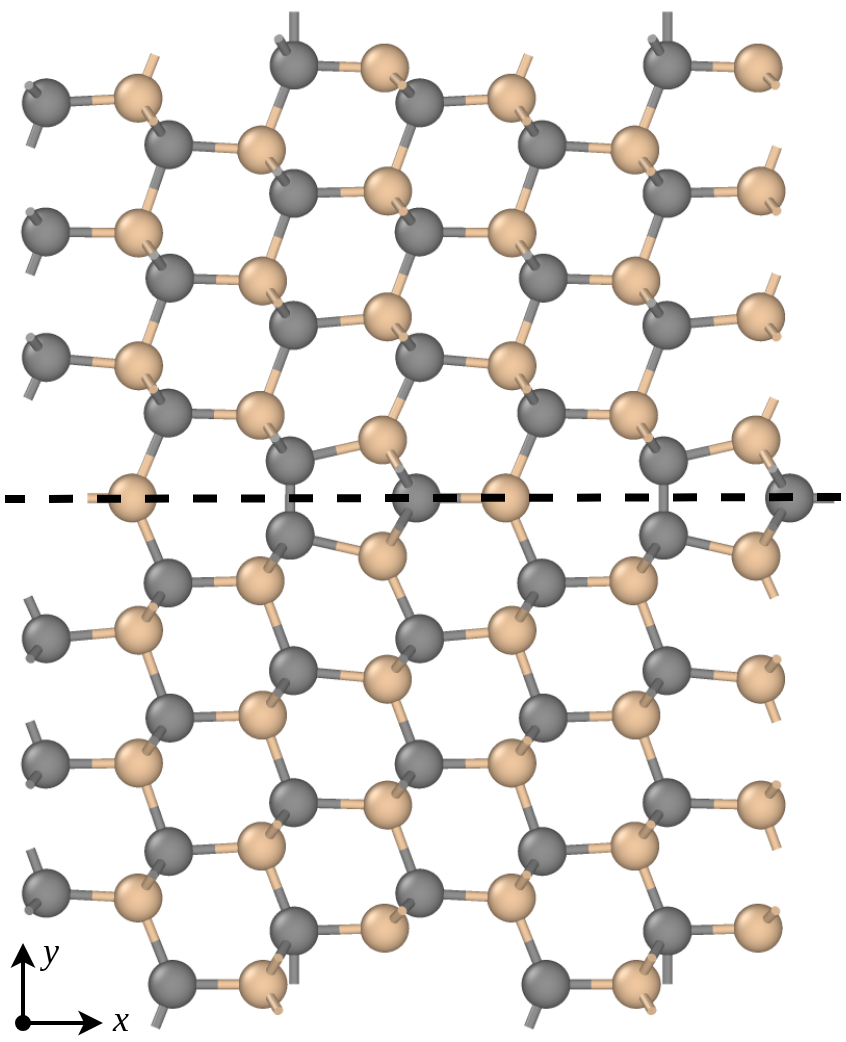}
 \caption{$p2'mm' (1 \times 2)$ model}
 \label{Fig:S3-112-2}
\end{subfigure}
\caption{(Color online) (\textbf{a}) Construction of the $\Sigma 3 \{ 112 \}$ GB via the superposition of two symmetric crystals along the $\{112\}$ plane, which results in the energetically unfavorable $p2'mm' (1 \times 1)$ model. (\textbf{b}) Type A of the $p2'mm' (1 \times 2)$ symmetric reconstruction model of the $\Sigma 3 \{ 112 \}$ GB. C atoms are gray and Si atoms are light brown.}
\end{figure}

A simple construction of the $\Sigma 3 \{ 112 \}$ GB via the superposition of two symmetric crystals along the $\{112\}$ plane results in the $p2'mm' (1 \times 1)$ model (\cref{Fig:S3-112-1}). However, this configuration is energetically unfavorable due to the presence of dangling bonds, floating bonds, and wrong bonds, and thus undergoes reconstruction~\cite{Kohyama1994, Tanaka2002}. Two reconstructed models have been proposed: (\textit{a}) the $p2'mm' (1 \times 2)$ symmetric model and (\textit{b}) the $pm (1 \times 2)$ asymmetric model. Each of these can exhibit two subtypes: \textit{Type A}, where Si atoms are reconstructed and C-C wrong bonds are formed, and \textit{Type B}, where C atoms are reconstructed and Si-Si wrong bonds are formed \cite{Kohyama1994, Tanaka2002}. Type A configurations are always more stable than Type B within both symmetric and asymmetric reconstructions \cite{Kohyama1994, Tanaka2002}. The energy difference between the Type A structures of the symmetric and asymmetric models is small \cite{Tanaka2002}, indicating that both can be realized, depending on the rigid-body translation conditions. However, because symmetric models do not involve rigid-body translations, they are easier to construct in MD simulations. Consequently, Type A of the $p2'mm' (1 \times 2)$ symmetric model (\cref{Fig:S3-112-2}) is often selected as the representative structure for the $\Sigma 3 \{ 112 \}$ GB \cite{Khalil2011}.

As mentioned earlier, the $\Sigma 9 \{122\}$ GB exhibits two polar and one non-polar interface. One polar interface, Type A, contains C-C wrong bonds, the other contains Si-Si wrong bonds, and the non-polar interface contains both types of wrong bonds. The formation energy of polar interface pairs is lower than that of non-polar interfaces due to reduced bond distortion energy \cite{Kohyama1991}, as each polar interface contains only one wrong bond type. Additionally, polar interfaces are always more stable than the non-polar interface \cite{Kohyama1991}. In this study, all three types of the $\Sigma 9$ GB are considered due to their structural simplicity. These boundaries do not require rigid body translations for their construction and can be derived from one another by interchanging the sublattices in either or both of the bicrystal's constituent grains.


\subsubsection{GB construction and diffusivity calculation}

Next, we describe the computational setup used to estimate Ag diffusivities in $\Sigma 3$ and $\Sigma 9$ GBs using MD simulations. All MD calculations performed in this work utilize the Large-scale Atomic/Molecular Massively Parallel Simulator (LAMMPS) software package \cite{Thompson2022}. A time step of 1 fs is utilized, and periodic boundary conditions (PBCs) are applied in all directions. The Open Visualization Tool (OVITO) software package \cite{Stukowski2010} is used for supercell visualization and analysis.

The analytic bond-order potential (ABOP) developed by Chen \textit{et al.} \cite{Chen2019} is employed to model the Si-C-Ag system. This potential has been validated against both experimental measurements and DFT calculations. In particular, it predicts a migration energy of 1.17 eV for Ag interstitial diffusion in bulk 3C-SiC, which closely matches the value obtained from spin-polarized DFT calculations (1.09 eV) \cite{Peng2019}. This agreement supports the suitability of the ABOP for simulating Ag diffusion in 3C-SiC.

Bicrystals are constructed as follows to model $\Sigma 3$ and $\Sigma 9$ GBs. For $\Sigma 3 \{112\} \langle 110 \rangle$, the axes of the upper grain are: $\mathbf{x}_u = [ \bar{1} 1 \bar{1} ]$, $\mathbf{y}_u = [011]$, and $\mathbf{z}_u = [2 1 \bar{1}]$ \cite{Kohyama1994}. For $\Sigma 9 \{ 122 \} \langle 110 \rangle$, the axes of the upper grain are: $\mathbf{x}_u = [\bar{4} \bar{1} 1]$, $\mathbf{y}_u = [011]$, and $\mathbf{z}_u = [\bar{1} 2 \bar{2}]$ \cite{Kohyama1991}. For both $\Sigma 3$ and $\Sigma 9$ GBs, the axes of the lower grain are obtained by $\mathbf{x}_l = \mathbf{x}_u$, $\mathbf{y}_l = - \mathbf{y}_u$, and $\mathbf{z}_l = - \mathbf{z}_u$. For both $\Sigma 3$ and $\Sigma 9$ GBs, the GB plane is the $xy$-plane. The supercell dimensions along the $x$-, $y$- and $z$- axes are: $l_x$ = 20 $p_x$, $l_y$ = 20 $p_y$, and $l_z$ = 60 {\AA}, where:
\begin{equation}
p_x = \sqrt{ h_x^2+ k_x^2 + l_x^2 } \ a, \quad p_y = \sqrt{ h_y^2+ k_y^2 + l_y^2 } \ a,
\end{equation}
with $\mathbf{x}_u = [h_x k_x l_x]$, $\mathbf{y}_u = [h_y k_y l_y]$, and $a$ the lattice constant. The ABOP predicted the 0~K lattice constant of 3C-SiC to be 4.359 {\AA}.

\begin{figure}[h!]
\centering
\begin{subfigure}{0.3\textwidth}
 \includegraphics[width=\textwidth]{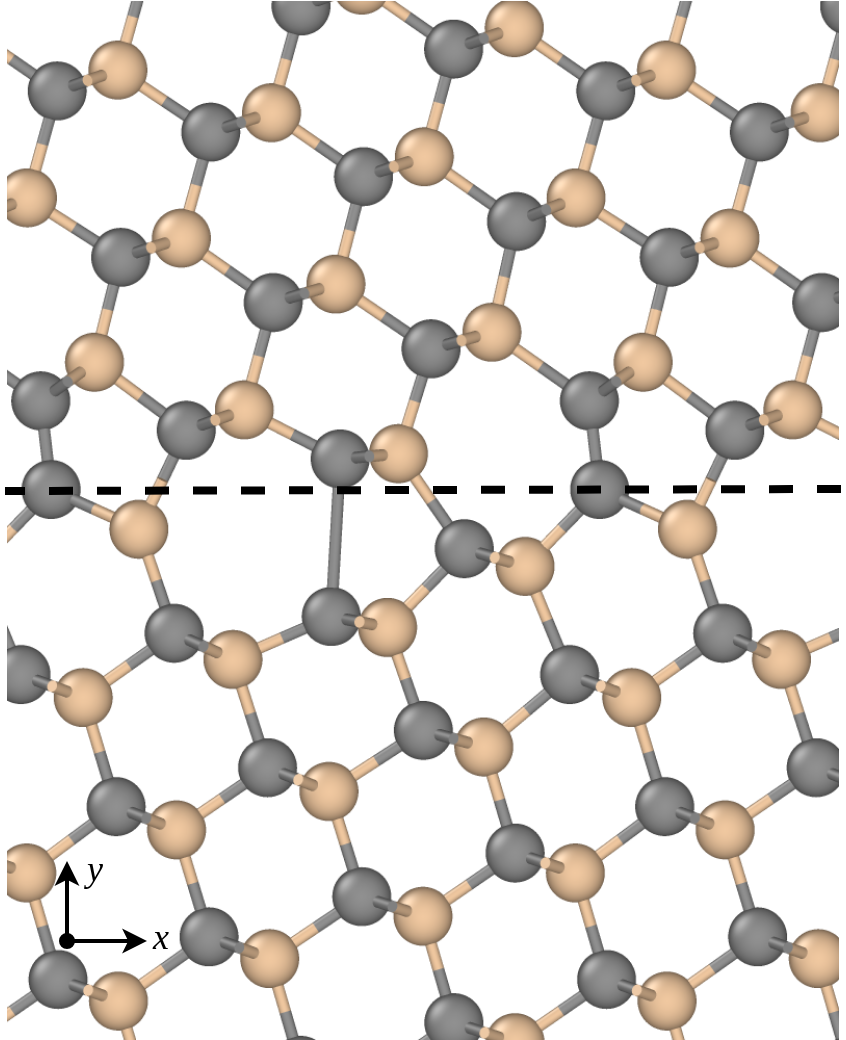}
 \caption{Type A: C-rich $\Sigma 9$ GB}
 \label{Fig:S9-A}
\end{subfigure}
\hfill
\begin{subfigure}{0.3\textwidth}
 \includegraphics[width=\textwidth]{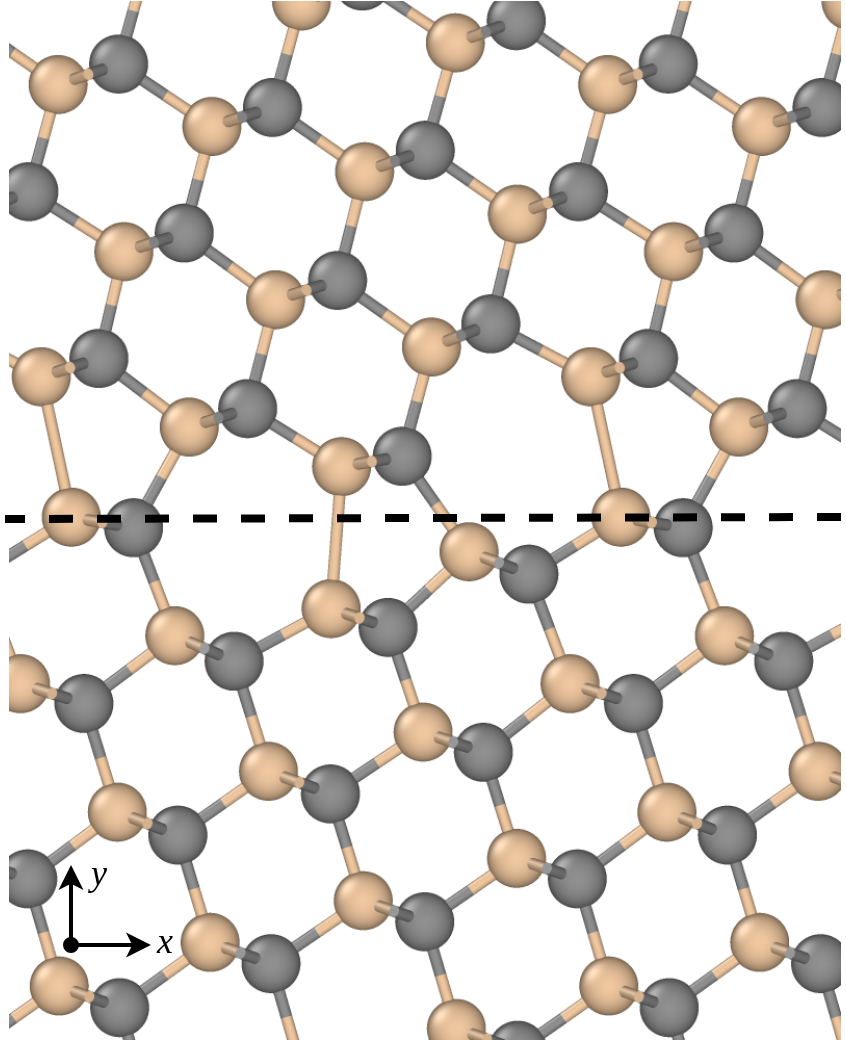}
 \caption{Type B: Si-rich $\Sigma 9$ GB}
 \label{Fig:S9-B}
\end{subfigure}
\hfill
\begin{subfigure}{0.3\textwidth}
 \includegraphics[width=\textwidth]{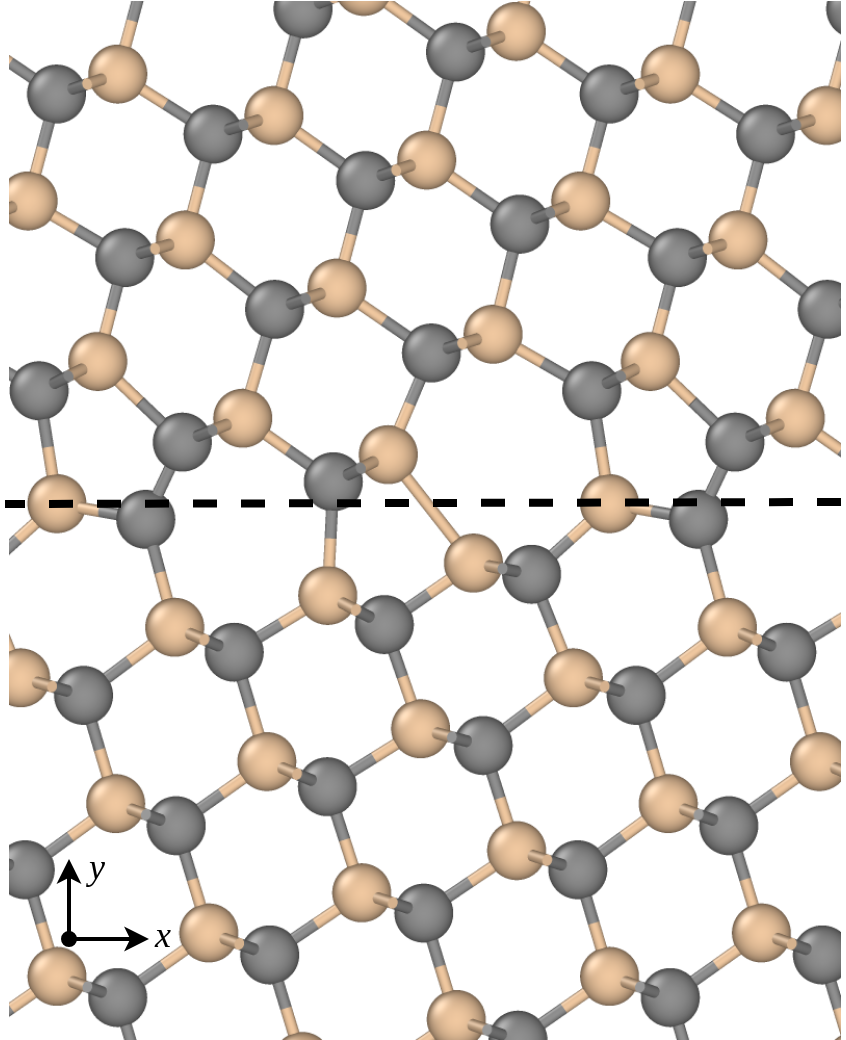}
 \caption{Type C: Non-polar $\Sigma 9$ GB}
 \label{Fig:S9-C}
\end{subfigure}
\caption{(Color online) (\textbf{a}) Polar variant of the $\Sigma 9$ GB that contains C-C wrong bonds, i.e., Type A. (\textbf{b}) Polar variant of the $\Sigma 9$ GB that contains Si-Si wrong bonds, i.e., Type B. (\textbf{c}) Non-polar variant of the $\Sigma 9$ GB that contains both types of wrong bonds, i.e., Type C. C atoms are gray and Si atoms are light brown.}
\label{Fig:S9}
\end{figure}

To determine the stable atomic configurations, energy minimization at 0~K was employed for each bicrystal. The resulting reconstructed structure of the $\Sigma 3 \{112\}$ GB following minimization is presented in \cref{Fig:S3-112-2}, while the various configurations of the $\Sigma 9$ GB are shown in \cref{Fig:S9}. The obtained structures of the $\Sigma 3$ and $\Sigma 9$ GBs are validated through comparison with configurations reported in prior studies based on tight-binding methods \cite{Kohyama1991,Kohyama1994,Tanaka2002}, MD simulations \cite{Kohler2002,Khalil2011}, and DFT calculations \cite{SynowczynskiDunn2020}.

The atomistic mechanism of Ag diffusion along the $\Sigma 3$ GB has been qualitatively discussed by Khalil \textit{et al.} \cite{Khalil2011}, although their quantitative results require a revision due to data processing errors. In contrast, the diffusion behavior in the $\Sigma 9$ GB remains largely unexplored. To investigate this, we inserted 1--10 Ag interstitials into the GB width of the various $\Sigma 9$ GB configurations, including atoms distributed among the rings, and atoms inserted randomly within the GB width. Here, the GB width is defined as the spatial region encompassing all atoms deviating from the cubic-diamond structure.

To obtain reliable Arrhenius fits for Ag diffusion in $\Sigma 9$ and $\Sigma 3$ GBs, good sampling statistics are required. Therefore, a concentration of 0.01 at.\% Ag is employed to ensure adequate sampling of diffusion events while maintaining dilute conditions. This level of doping remains lower than the concentrations adopted in prior studies~\cite{Jiang2021, Aagesen2022}, which examined Ag diffusion in $\Sigma 5$ and HAGBs.

After inserting the Ag atoms, the bicrystals are allowed to evolve at zero pressure and $T$ = 1000--2000~K for 10 ns under the Nose-Hoover thermostat-barostat, where the mean squared displacements (MSDs) are calculated. 2-D diffusion is assumed for all GBs. The diffusivity, $D$, at each temperature is calculated according to:
\begin{equation}
D = \lim_{t \rightarrow \infty} \frac{\text{MSD}(t)}{2d \cdot t} = \lim_{t \rightarrow \infty} \frac{1}{2d \cdot t} \frac{ \sum_i \left| \mathbf{r}_{i}(t) - \mathbf{r}_i(0) \right|^2 }{ N },
\label{Eq:D1}
\end{equation}
where $d$ is the diffusion dimensionality ($d=2$), $t$ is time, and $N$ is the number of Ag atoms. Then, all $D$ values are fitted to an Arrhenius relation:
\begin{equation}
D(T) = D_0 \ \text{exp}\left( - \frac{Q}{k T} \right),
\label{Eq:D2}
\end{equation}
where $D_0$ is the diffusion pre-factor, $Q$ is the activation energy, and $k$ is Boltzmann's constant.


\subsection{Bayesian calibration}

To characterize the effective diffusion parameters, $D_{0,\text{eff}}$ and $Q_\text{eff}$, of Ag in 3C‑SiC from literature data, we perform a Bayesian inference on the set of Ag diffusivity data derived from integral release and ion implantation experiments, as compiled by Ko \textit{et al.}~\cite{Ko2016} (\cref{Tab:LitData}). The objective is to obtain a joint posterior distribution for the Arrhenius pair that captures both parameter uncertainty and correlation structure, grounded in physical prior beliefs. The diffusion coefficient is assumed to follow the standard Arrhenius form:
\begin{equation}
\ln D_\text{eff}(T) = \ln D_{0,\text{eff}} - \frac{Q_\text{eff}}{kT}.
\end{equation}
The observational data consist of 12 experimentally reported pairs of $(D_{0,\text{obs}}, Q_\text{obs})$ values, which we log-transform to obtain $(\ln D_{0,\text{obs}}, Q_\text{obs})$ for direct use in regression.

\begin{table}[htbp]
\centering
\caption{Summary of experimental values for effective diffusion parameters of Ag in 3C-SiC, compiled by Ko \textit{et al.} \cite{Ko2016}.}
\begin{tabular}{S[table-format=1.2e-1] S[table-format=1.2]}
\hline
{$D_{0,\text{eff}}$ [m$^2$/s]} & {$Q_{\text{eff}}$ [eV]} \\
\hline
6.8e-9 & 2.21 \\
9.6e-6 & 4.24 \\
3.6e-9 & 2.23 \\
6.8e-11 & 1.84 \\
1.14e-13 & 1.13 \\
4.3e-12 & 2.50 \\
4.5e-9 & 2.26 \\
6.8e-11 & 1.84 \\
5.0e-10 & 1.89 \\
3.5e-10 & 2.21 \\
2.4e-9 & 3.43 \\
1.04e-12 & 1.84 \\
\hline
\end{tabular}
\label{Tab:LitData}
\end{table}

We carry out posterior inference using Hamiltonian Monte Carlo (HMC) sampling~\cite{HMC1,HMC2}, implemented via the \verb|PyMC| probabilistic programming library~\cite{PyMC}. HMC is a Markov Chain Monte Carlo (MCMC) method that uses Hamiltonian dynamics to generate distant, informed moves, overcoming the inefficiency of slow random-walk algorithms in traditional MCMC~\cite{Neal2011}. It improves over traditional MCMC methods by using gradient information to explore the parameter space more efficiently. These gradients are computed through automatic differentiation, which applies the chain rule across code operations to yield exact derivatives. This capability is essential for HMC's performance, as accurate gradients are required to simulate Hamiltonian dynamics. To enhance convergence and eliminate the need for manual tuning of trajectory lengths, we employ the No-U-Turn Sampler (NUTS)~\cite{NUTS}, an adaptive variant of HMC.

Each observed pair $(\ln D_{0,\text{obs}}, Q_{\text{obs}})$ is modeled as a noisy realization of a latent bivariate Gaussian distribution with unknown mean vector and covariance structure:
\begin{equation}
\begin{bmatrix}
\ln D_{0,\text{obs}} \\
Q_{\text{obs}}
\end{bmatrix}
\sim \mathcal{N}
\left(
\begin{bmatrix}
\mu_{\ln D_0} \\
\mu_Q
\end{bmatrix},
\boldsymbol{\Sigma}
\right),
\end{equation}
where $\Sigma \in \mathbb{R}^{2\times2}$ is the covariance matrix capturing both individual uncertainties and correlation between the parameters.

The mean vector $\mu = [\mu_{\ln D_0}, \mu_Q]$ is assigned a bivariate normal prior:
\begin{equation}
\boldsymbol{\mu} \sim \mathcal{N} \left(
\begin{bmatrix}
\bar{\mu}_{\ln D_0} \\
\bar{\mu}_Q
\end{bmatrix},
\mathrm{diag}(\sigma_{\ln D_0}^2, \sigma_Q^2)
\right),
\end{equation}
where $\bar{\mu}_{\ln D_0}$ and $\bar{\mu}_Q$ are set to the empirical means of the input literature values. We assessed the sensitivity of the posterior to the prior mean vector $\bar{\mu} = [\bar{\mu}_{\ln D_0}, \bar{\mu}_Q]$ by testing significantly different values, and found that the resulting posterior distributions remained nearly unchanged, indicating that the inference is robust to the choice of prior location due to the dominance of the likelihood. The prior standard deviations are chosen to reflect physically reasonable uncertainty: $\sigma_{\ln D_0} = 3 \ln 10$, representing three orders of magnitude uncertainty in $D_0$, and $\sigma_Q = 3$~eV. Intuitively, this uncertainty means that there is 68\% probability that the true mean of the study‑to‑study distribution lies within $\boldsymbol{\mu} + \boldsymbol{\sigma}$. Larger deviations are allowed but get exponentially less likely. This effectively represents uninformative priors.

The covariance matrix $\Sigma$ is constructed via a Cholesky decomposition \cite{McClarren2018} to ensure positive definiteness. Specifically, we define the marginal standard deviations as:
\begin{equation}
\boldsymbol{\sigma} = [\sigma_{\ln D_0}, \sigma_Q] \sim \mathrm{HalfNormal}(\text{scale} = [3 \ln 10, 3]).
\end{equation}

To model the correlation between the Arrhenius parameters, we assign a Lewandowski-Kurowicka-Joe (LKJ) prior~\cite{LKJ} to the correlation matrix $\mathbf{R} \in \mathbb{R}^{2\times2}$:
\begin{equation}
\textbf{R} \sim \mathrm{LKJ}(\eta = 2),
\end{equation}
which defines a probability distribution over all valid correlation matrices. The concentration parameter $\eta$ controls the strength of the prior belief about the correlation structure. When $\eta = 1$, all correlation coefficients, $\rho$, are equally likely (uniform prior). As $\eta > 1$, the prior increasingly favors correlation coefficients near zero. We set $\eta = 2$ to weakly prefer moderate correlations (i.e., discourage extreme values such as $|\rho| \approx 1$) without ruling them out. This allows strong correlations to emerge if supported by the data, while avoiding overfitting when the evidence is weak.

The need for such a correlation structure is physically motivated. Consider the Arrhenius expression for diffusivity:
\begin{equation}
D(T) = D_0 \exp\left(-\frac{Q}{kT}\right).
\label{Eq:Arrhenius}
\end{equation}
If we assume that $D(T)$ is fixed around a given temperature and that both $D_0$ and $Q$ are uncertain, then their variations must be coupled. Taking the total differential of \cref{Eq:Arrhenius} and enforcing $\delta D = 0$, we obtain:
\begin{equation}
\delta D = \exp\left(-\frac{Q}{kT}\right) \left( \delta D_0 - \frac{D_0}{kT} \, \delta Q \right) = 0,
\end{equation}
which gives:
\begin{equation}
\frac{\delta D_0}{D_0} = \frac{\delta Q}{kT}.
\label{Eq:Coupling}
\end{equation}

\cref{Eq:Coupling} shows that an increase in activation energy $Q$ must be offset by an increase in $D_0$ to preserve a constant value of $D(T)$. This induces an intrinsic positive correlation between $\ln D_0$ and $Q$. The LKJ prior with $\eta = 2$ flexibly captures this physically meaningful correlation.

Finally, the full covariance matrix is then given by:
\begin{equation}
\boldsymbol{\Sigma} = \mathrm{diag}(\boldsymbol{\sigma}) \cdot \mathbf{R} \cdot \mathrm{diag}(\boldsymbol{\sigma}).
\end{equation}

We perform inference using four independent NUTS chains, each run for 4000 posterior draws after 1000 warm-up steps. The target acceptance rate is set to 0.9 to improve sampling robustness and avoid divergences. Posterior marginals and joint samples are visualized using kernel density estimation (KDE) \cite{Smith2014} and posterior scatter overlays, respectively. These diagnostics inspect the adequacy of the model and the nature of the posterior.


\subsection{Local sensitivity analysis}

To quantify the influence of individual Arrhenius parameters on the effective diffusivity behavior, we perform a local sensitivity analysis. This analysis evaluates how variations in the activation energy \( Q_i \) and pre-exponential factor \( D_{0,i} \) for each diffusion pathway, including the bulk and distinct GB types, affect the quantities of interest (QoIs), namely $Q_{\text{eff}}$ and $D_{0,\text{eff}}$. The nominal parameter vector is defined as:
\begin{equation}
\mathbf{x^0} = [Q_i, D_{0,i}], \quad i \in \{\text{Bulk}, \Sigma3, \dots\}.
\end{equation}

A small finite step size $h_i = 10^{-6} x_i^0$ is applied to each parameter $x_i^0$ using a central finite-difference scheme \cite{McClarren2018}. The resulting central finite-difference derivative quantifies the local sensitivity of QoIs to small perturbations in the model input parameters. Sensitivity indices are then computed using the following expression \cite{Saltelli2004}:
\begin{equation}
S_{j,i} = \frac{x_i^0}{y_j^0} \frac{y_j(x_i^0 + h_i) - y_j(x_i^0 - h_i)}{2h_i}, \quad \mathbf{y} = [Q_{\text{eff}}, D_{0,\text{eff}}],
\label{Eq:LocalSi}
\end{equation}
where $S_{j,i}$ is the sensitivity index of output $y_j$ to input $x_i$, $x_i^0$ and $y_j^0$ are the nominal values of $x_i$ and $y_j$, and $y_j$ is either $Q_\text{eff}$ or $D_\text{0,eff}$. The sensitivity index, as defined in \cref{Eq:LocalSi}, is local and does not require knowledge of the input parameter uncertainties. A local sensitivity analysis is chosen instead of a global one because not all reported values of $D_{0, i}$ and $Q_i$ include associated standard deviations. Note that the sensitivity indices are normalized with respect to both the parameter values and the output quantities, making them dimensionless and suitable for cross-comparison.


\section{Results}

\subsection{Atomistic mechanism of Ag diffusion}

\begin{figure}[b!]
\centering
\begin{subfigure}[b]{0.50\textwidth}
  \centering
  \includegraphics[height=4.5cm]{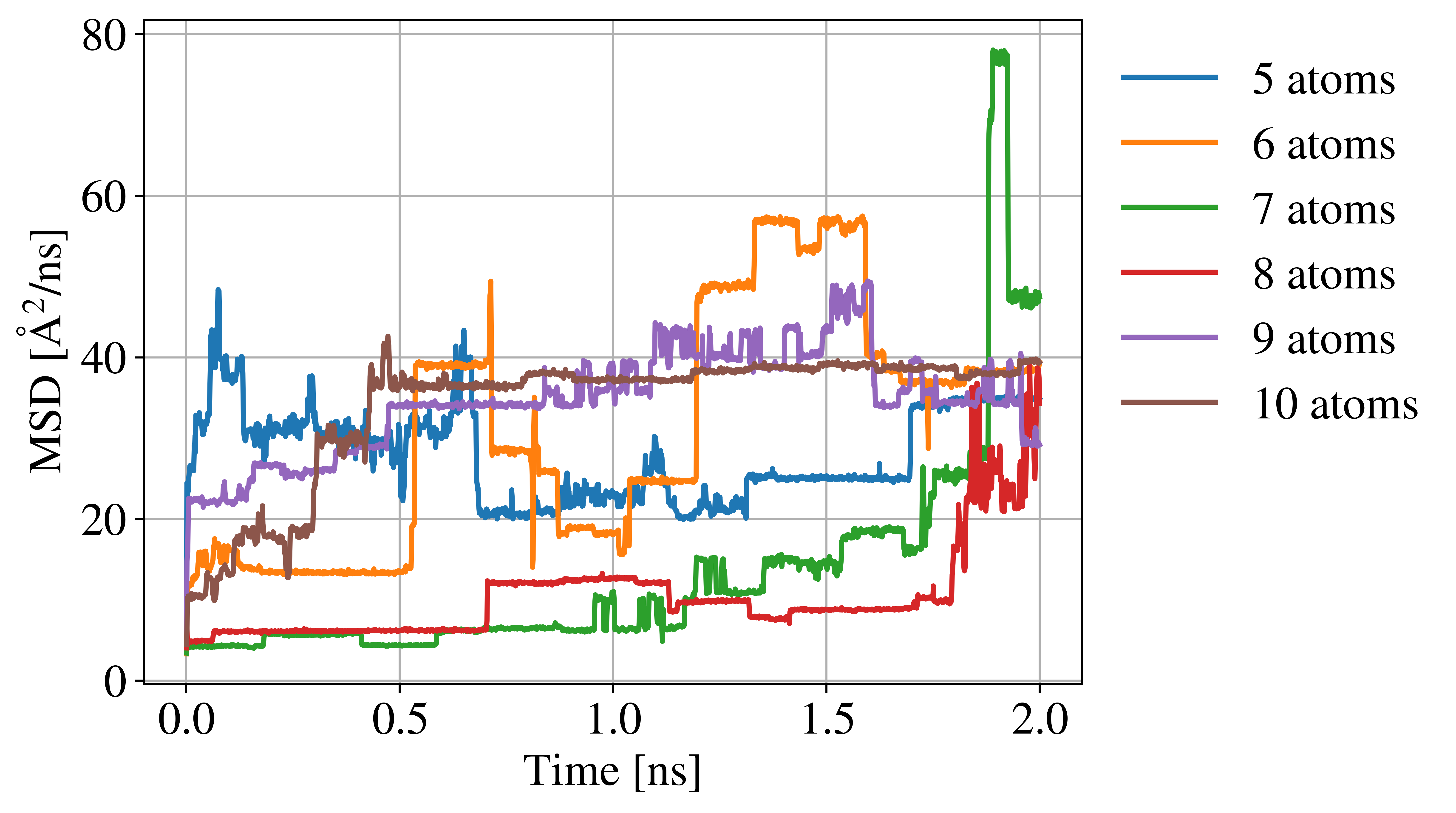}
  \caption{}
  \label{Fig:InitialConfig}
\end{subfigure}
\hspace{1em}
\begin{subfigure}[b]{0.45\textwidth}
  \centering
  \includegraphics[height=4.5cm]{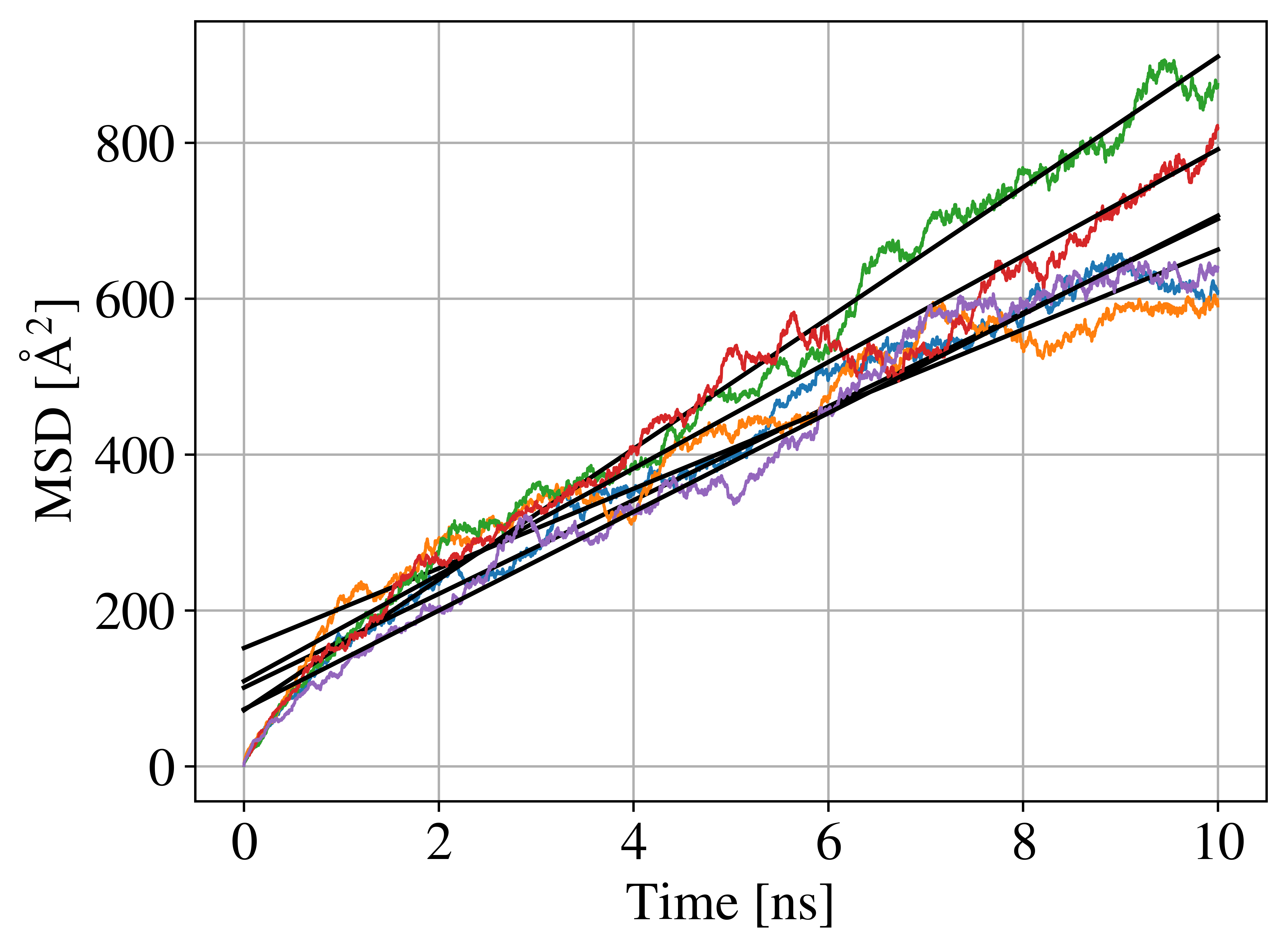}
  \caption{}
  \label{Fig:0.1Ag}
\end{subfigure}
\caption{(Color online) (\textbf{a}) MSD versus time plot for C-rich $\Sigma 9$ GB at 1400~K with 5--10 Ag atoms. (\textbf{b}) MSD versus time plot for Si-rich $\Sigma 9$ GB at 1500~K with 0.1 at.\% Ag atoms. All Ag atoms have been inserted randomly in the GB width.}
\end{figure}

We begin by examining the atomistic mechanisms underlying Ag diffusion in $\Sigma 9$ GBs containing 5--10 Ag atoms. Several characteristics can be inferred from the analysis of MSD versus time plots combined with the visual inspection of Ag atom trajectories. An example MSD versus time curve for the C-rich $\Sigma 9$ GB at 1400~K is shown in \cref{Fig:InitialConfig}. 


Ag atoms exhibit a strong preference for occupying 7-atom ring structures over 5-atom ring structures, with enhanced mobility along the direction aligned with the 7-atom rings (see \cref{Fig:S9}). The MSD of Ag atoms displays significant variability across simulations conducted under nearly identical macroscopic conditions. This sensitivity arises from differences in initial atomic configurations. Specifically, placing Ag atoms inside or near favorable ring structures (particularly 7-atom rings) can markedly enhance diffusion.

Following insertion, Ag atoms rapidly form transient clusters that dominate net mass transport within the GB region. These clusters may intermittently dissociate due to thermal agitation, allowing individual atoms to occupy substitutional sites. Once in such sites, Ag atoms often become immobilized, thereby reducing the overall diffusivity. Nevertheless, thermally activated escape events can occasionally liberate these trapped atoms, as illustrated in \cref{Fig:InitialConfig}, which shows flat MSD segments interrupted by sudden jumps. At low and intermediate temperatures, thermal energy is insufficient to overcome trapping barriers for dilute concentrations of silver, suppressing diffusion. Conversely, higher temperatures and/or higher Ag concentrations facilitate escape from these traps, resulting in smoother MSD curves. These observations are the same for all variants of the $\Sigma 9$ GB. Overall, simulations of $\Sigma 9$ GBs containing 5--10 Ag atoms indicate that Ag diffusion in these GBs is governed by cluster-mediated motion, strongly influenced by local ring structures and the distinct dynamics of substitutional versus interstitial atoms. This behavior is consistent with findings in bulk SiC reported by~\cite{Chen2019, Jiang2021}.

\begin{figure}[h!]
\centering
\includegraphics[width=0.8\textwidth]{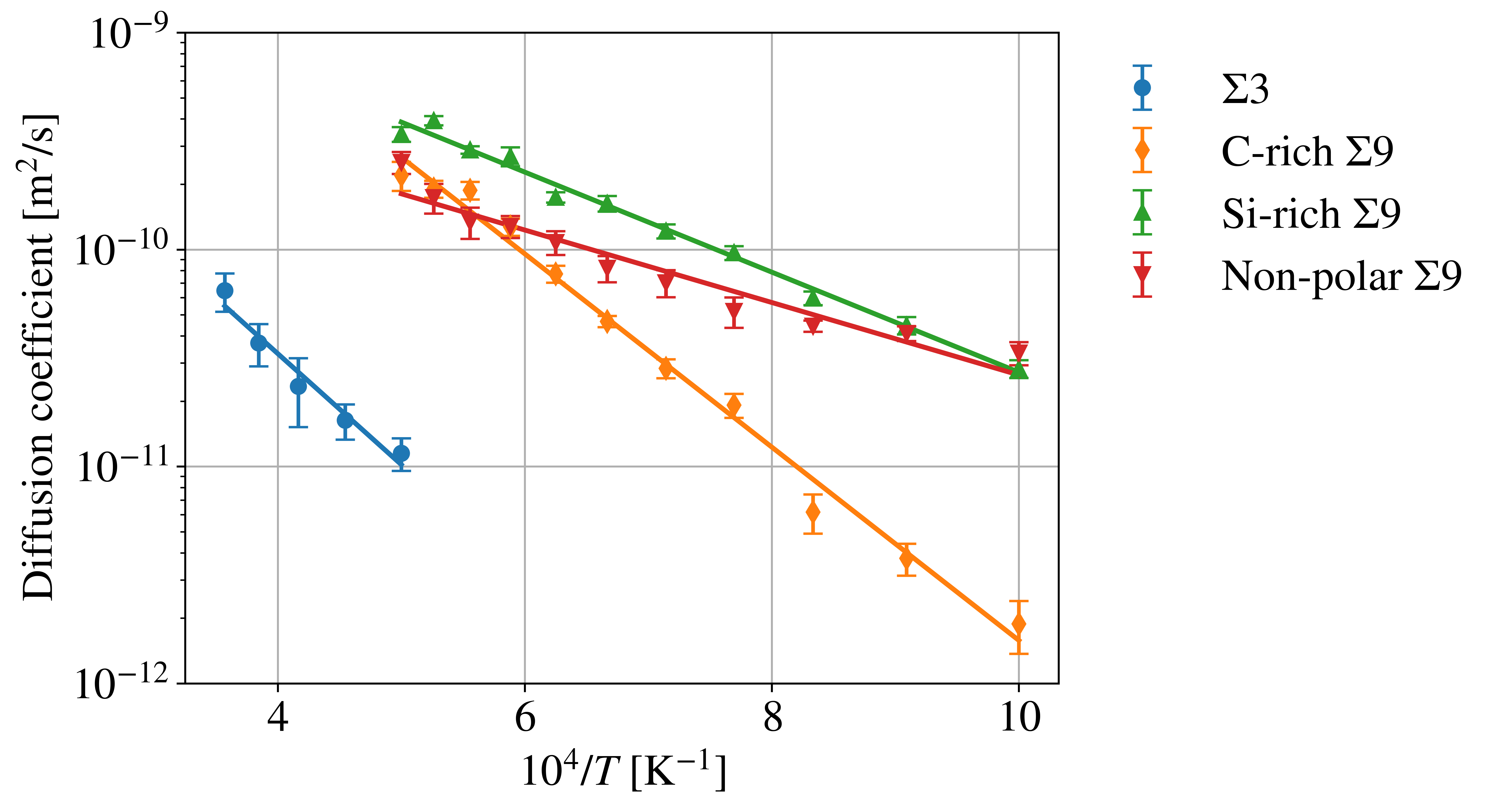}
\caption{Arrhenius plots of Ag diffusion in $\Sigma 3$ GBs ($D_0 = 3.69 \times 10^{-9}$ m$^2$/s, $Q$ = 1.02 eV), C-rich $\Sigma 9$ GBs ($D_0 = 4.50 \times 10^{-8}$ m$^2$/s, $Q$ = 0.88 eV), Si-rich $\Sigma 9$ GBs ($D_0 = 5.51 \times 10^{-9}$ m$^2$/s, $Q$ = 0.46 eV), and non-polar $\Sigma 9$ GBs ($D_0 = 1.24 \times 10^{-9}$ m$^2$/s, $Q$ = 0.33 eV). All supercells contain 0.1 at.\% Ag, which corresponds to 105 Ag atoms out of 105,705 total atoms in the $\Sigma 3$ supercells and 194 Ag atoms out of 194,594 total atoms in the $\Sigma 9$ supercells. $\Sigma 3$ diffusivity is fitted at 2000--2800~K, whereas $\Sigma 9$ diffusivity is fitted at 1000--2000~K. Linear fits yield effective activation energies and pre-exponential factors. Data points represent diffusivities averaged across five independent simulations. Error bars represent standard error.}
\label{Fig:Sigma}
\end{figure}


After examining $\Sigma 9$ supercells with dilute Ag concentrations to infer the atomistic diffusion mechanisms, we next obtain statistically significant diffusivity predictions by analyzing $\Sigma 3$ and $\Sigma 9$ supercells containing 0.1 at.\% Ag within the GB width (see \cref{Fig:0.1Ag}). This concentration corresponds to 105 Ag atoms out of 105,705 total atoms in the $\Sigma 3$ supercells and 194 Ag atoms out of 194,594 total atoms in the $\Sigma 9$ supercells. As shown in \cref{Fig:Sigma}, a clear hierarchy of diffusivity emerges across the temperature range $T$ = 1000--2000~K. Specifically, Ag diffusion is fastest in the non-polar $\Sigma 9$ GB, followed by the Si-rich GB, and slowest in the C-rich GB. This trend is reflected in the activation energies, with $Q$ = 0.88 eV for the C-rich GB, 0.46 eV for the Si-rich GB, and 0.33 eV for the non-polar GB. While the non-polar interface offers the highest Ag diffusivity, it is also the least thermodynamically stable \cite{Kohyama1991} and exhibits greater uncertainty in the diffusion data. In contrast, both polar variants (C-rich and Si-rich) are more stable and thus more likely to occur in realistic microstructures. For the purpose of modeling Ag transport, the C-rich variant is chosen to conservatively characterize the $\Sigma 9$ GB.

We found that the $\Sigma 3$ GB exhibits considerable noise and scatter in their MSD data, making it difficult to extract a reliable Arrhenius fit at $T$ = 1000--2000 K. This is because, at low temperatures, events with a high activation energy are rare, which require prohibitively long MD simulations to observe a sufficient number of these events to get statistically meaningful MSD plots. At elevated temperatures, increased kinetic energy accelerates diffusion, allowing sufficient sampling of rare events within feasible simulation times. To address this sampling problem, we conduct the $\Sigma 3$ GB simulations at $T$ = 2000--2800~K under the assumption that the same diffusion mechanisms operate across the entire temperature range. Note that high-temperature dynamics may activate multiple diffusion pathways, and as a result, the extracted activation energy represents an effective average over various mechanisms. As shown in \cref{Fig:Sigma}, although the standard error remains non-negligible, an Arrhenius fit becomes feasible and yields an activation energy of $Q$ = 1.02 eV, higher than that predicted for the $\Sigma 9$ GB.


\subsection{Bayesian calibration}

In this section, we perform Bayesian inference to jointly estimate the mean and correlation of the Arrhenius parameters from literature data. The posterior distribution of the mean vector, $\boldsymbol{\mu}$, is summarized in \cref{Tab:Posterior}, which reports the posterior means, standard deviations, and 95\% highest density intervals (HDIs) for both parameters. The posterior mean of $D_{0,\text{eff}}$ is found to be approximately $3.21 \times 10^{-10}$ m$^2$/s, while that of $Q_\text{eff}$ is 2.31 eV. The potential scale reduction factor, $\hat{R}$, \cite{Gelman1992} evaluates whether multiple sampling chains have converged to the same posterior distribution, with values near 1 indicating stable and well-mixed inference. For all monitored parameters, $\hat{R} = 1$ to six decimal places, indicating excellent convergence and supporting the reliability of the posterior estimates. The quality of the posterior sampling was assessed using effective sample size (ESS) diagnostics \cite{Vehtarh2021} for both bulk and tail estimates. For $\mu_{\ln D_{0,\text{eff}}}$, the bulk and tail ESS values are 6603 and 7885, respectively, while for $\mu_{Q_\text{eff}}$, the corresponding values were 7113 and 7582. These high ESS values indicate that the Markov chains efficiently explored both the central region and the tails of the posterior distributions, yielding stable and reliable estimates of the posterior mean and credible intervals.

\begin{table}[h!]
\centering
\caption{Posterior summary of Arrhenius parameters. HDI refers to the highest density interval.}
\label{Tab:Posterior}
\begin{tabular}{lccc}
\hline
Parameter & Posterior mean & Standard deviation & 95\% HDI \\
\hline
$\ln D_{0,\text{eff}}$ [ln m$^2$/s] & $-21.857$ & 1.429 & [$-24.654$, $-18.991$] \\
$D_{0,\text{eff}}$ [m$^2$/s] & $3.21 \times 10^{-10}$ & -- & [$1.96 \times 10^{-11}$, $5.65 \times 10^{-9}$] \\
$Q_\text{eff}$ [eV] & 2.31 & 0.25 & [1.77, 2.78] \\
\hline
\end{tabular}
\end{table}

\begin{figure}[h!]
\centering
\includegraphics[width=0.6\textwidth]{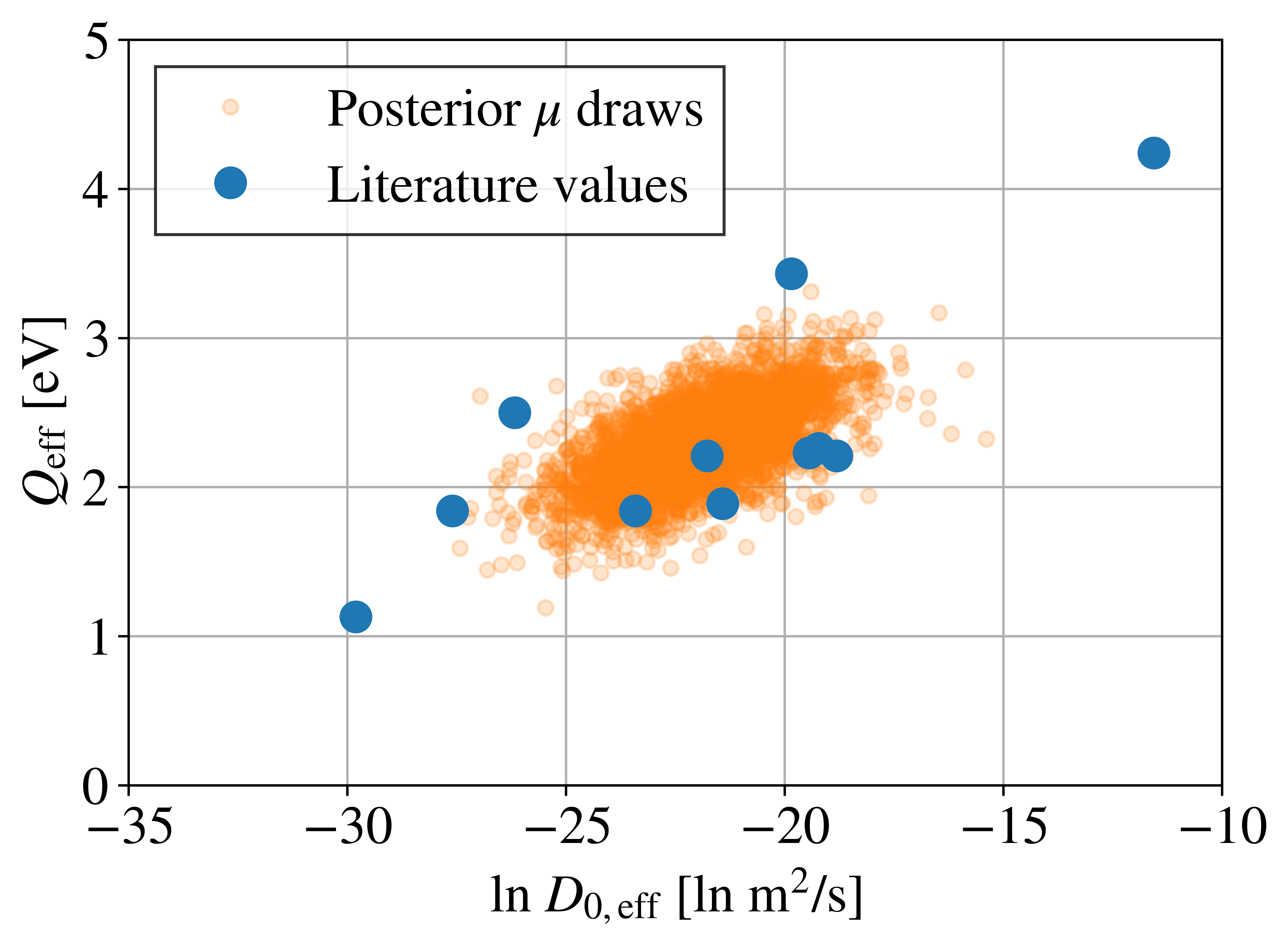}
\caption{(Color online) Joint posterior samples of the model parameters $\boldsymbol{\mu} = [\mu_{\ln D_0},\, \mu_Q]$ inferred from literature data using a Bayesian multivariate normal model. Orange points represent posterior draws of the bivariate mean vector, and blue circles correspond to the original experimental data points from the literature.}
\label{Fig:JointPosterior}
\end{figure}

\cref{Fig:JointPosterior} displays the joint posterior samples of the model parameters $\boldsymbol{\mu} = [\mu_{\ln D_0},\, \mu_Q]$, which represent the location parameters of the bivariate normal distribution fitted to the observed $(\ln D_0, Q)$ pairs. The posterior samples form a concentrated cloud near the centroid of the observed data, reflecting uncertainty in the population-level parameters rather than individual measurements. Notably, the posterior exhibits a positive correlation between $\mu_{\ln D_0}$ and $\mu_Q$, a physically meaningful relationship that is naturally captured by the LKJ prior and the data-driven Bayesian inference. Specifically, the inferred posterior correlation coefficient between $\mu_{\ln D_{0,\text{eff}}}$ and $\mu_{Q_\text{eff}}$ is $\rho = 0.818$, with a 95\% highest density interval (HDI) of [0.397, 1.000]. 

\begin{figure}[h]
\centering
\begin{subfigure}[b]{0.48\textwidth}
 \includegraphics[width=\textwidth]{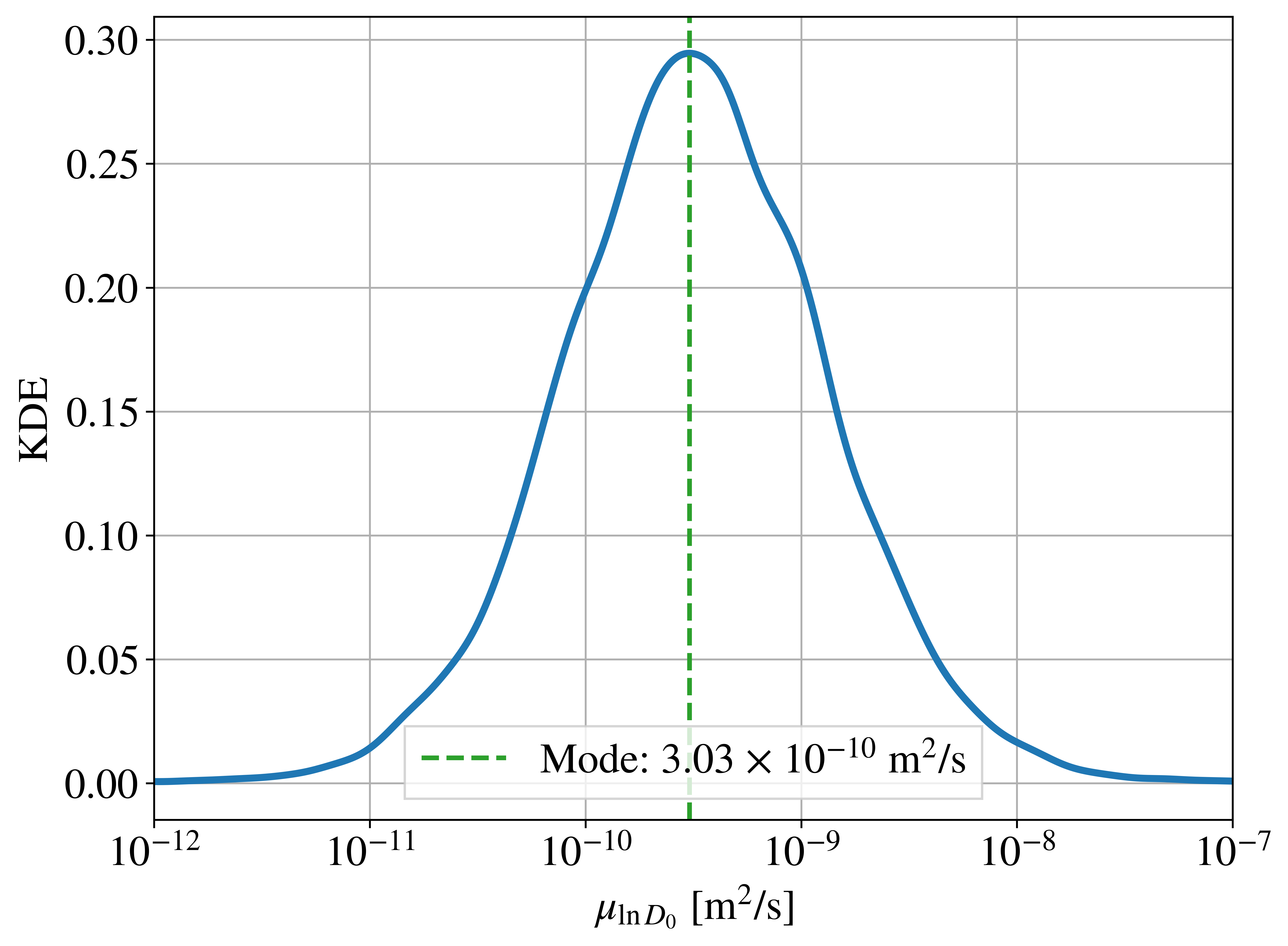}
 \caption{}
 \label{Fig:KDE_D0}
\end{subfigure}
\hfill
\begin{subfigure}[b]{0.48\textwidth}
 \includegraphics[width=\textwidth]{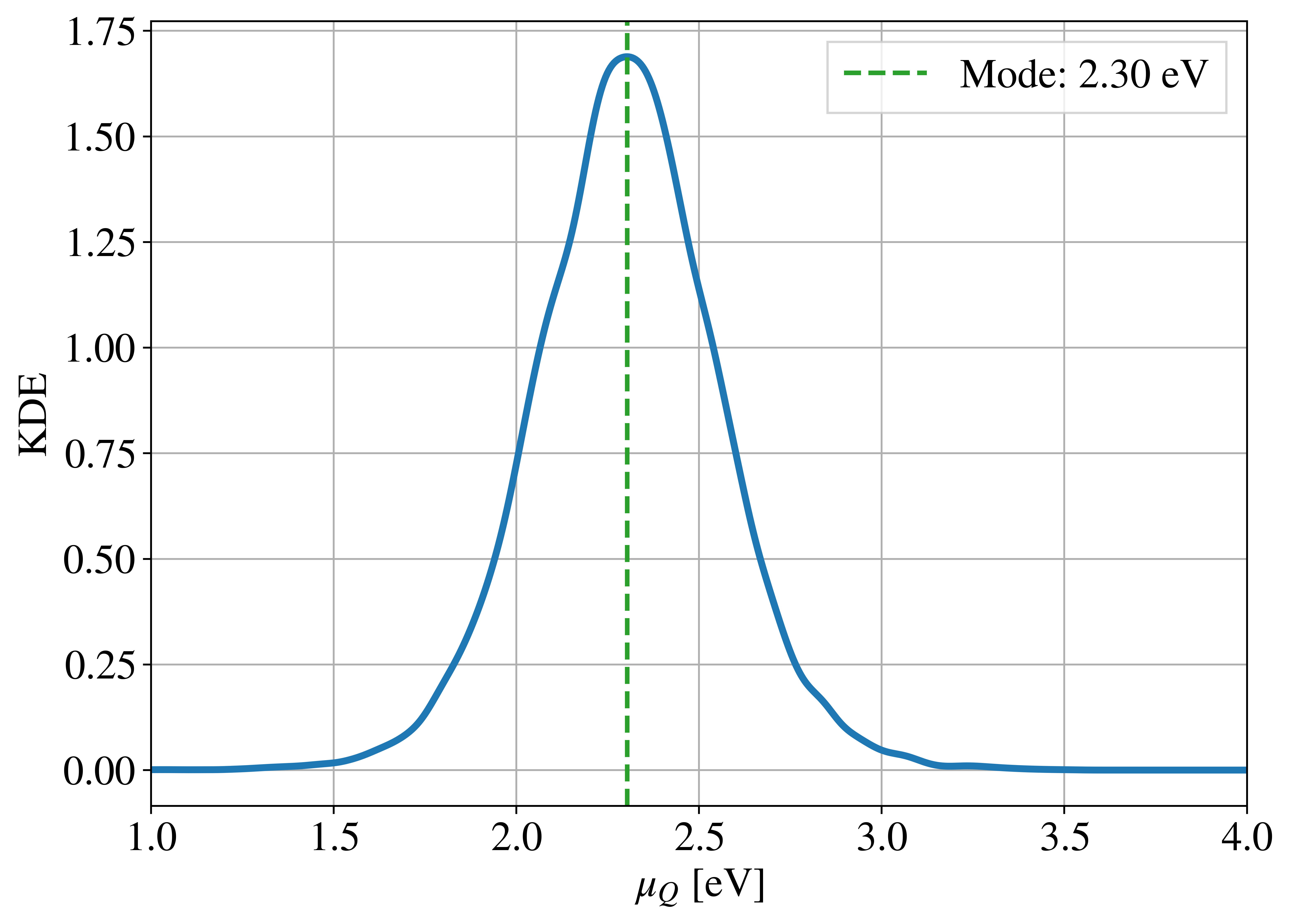}
 \caption{}
 \label{Fig:KDE_Q}
\end{subfigure}
\caption{Posterior marginal probability density functions (PDFs) of (\textbf{a}) $\mu_{\ln D_{0}}$ and (\textbf{b}) $\mu_{Q}$, estimated via kernel density estimation (KDE). The mode of $\mu_{\ln D_{0}}$ corresponds to $3.03 \times 10^{-10}$~m$^2$/s, and the mode of $\mu_Q = 2.30$~eV.}
\end{figure}

The marginal posterior distributions are visualized in \cref{Fig:KDE_D0,Fig:KDE_Q}, showing smoothed kernel density estimates (KDEs) for $\mu_{\ln D_{0,\text{eff}}}$ and $\mu_{Q_\text{eff}}$, respectively. These distributions reveal concentrated regions of high posterior density with limited tail behavior, indicating that the data are informative in constraining both parameters. The mode of the posterior for $\mu_{D_{0,\text{eff}}}$ is approximately $3.03 \times 10^{-10}$ m$^2$/s, and the mode for $\mu_{Q_\text{eff}}$ is near 2.30 eV. These modes are close to the previously estimated means (\cref{Tab:Posterior}), consistent with the central tendencies observed in the literature data. This strong, statistically significant positive correlation reflects the compensation effect \cite{Watanabe1990} commonly observed in diffusion data, whereby higher activation energies are associated with larger prefactors to maintain consistent diffusivity at relevant temperatures.

Overall, the Bayesian inference framework yields a robust estimation of the Arrhenius parameters of Ag diffusion from experimental results, with quantified uncertainty and correlated posteriors that reflect both physical intuition and data constraints.


\subsection{Effective diffusivity model}

\begin{figure}[h]
\centering
\includegraphics[width=0.7\textwidth]{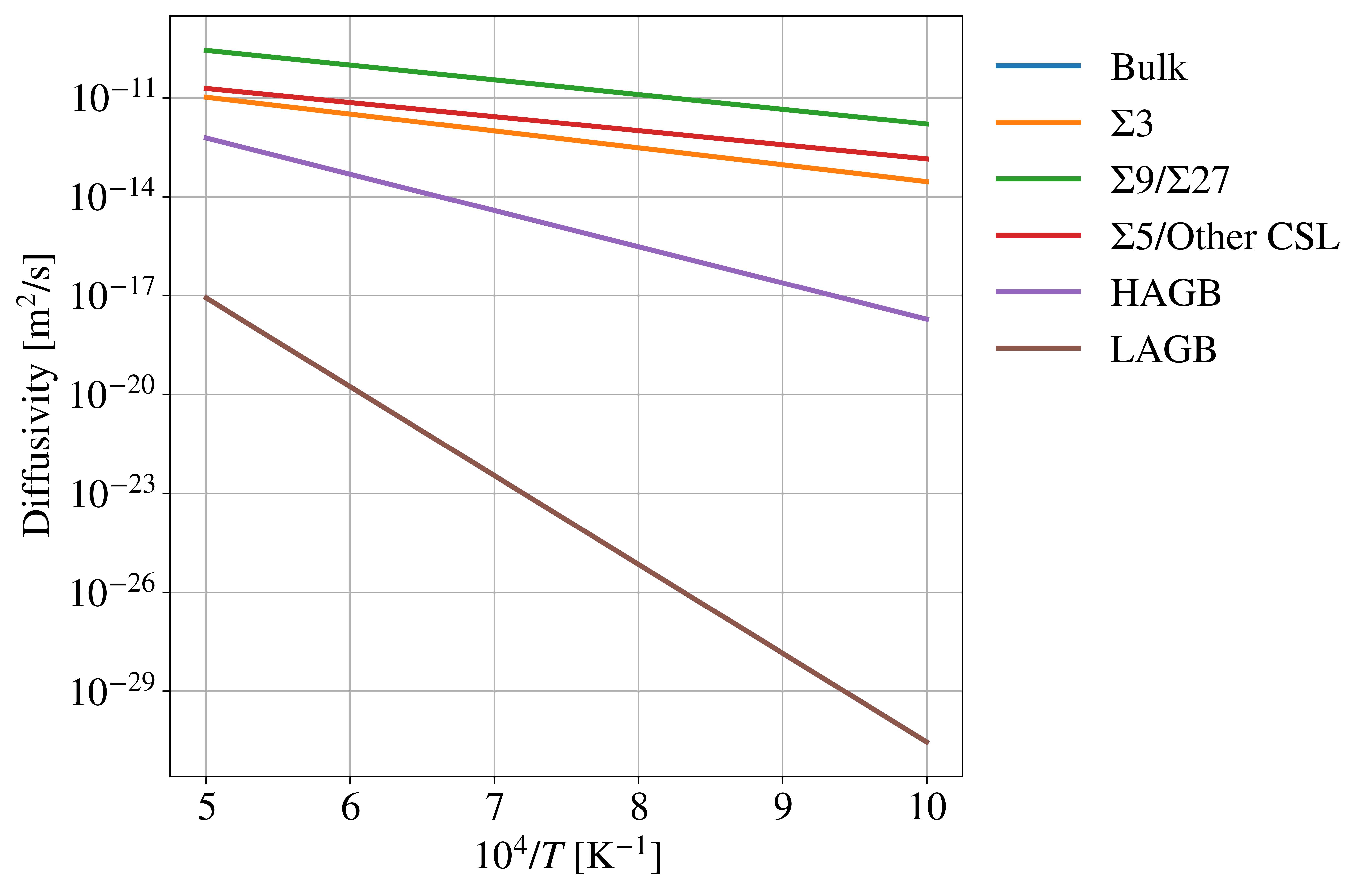}
\caption{(Color online) Arrhenius plot of the diffusivities of all considered GB types, as well as the bulk. The LAGB and bulk diffusivities coincide because they are assumed to have the same diffusivity.}
\label{Fig:ArrheniusAll}
\end{figure}

After estimating the most probable values of $D_{0,\text{eff}}$ and $Q_\text{eff}$, we next assess how well our proposed effective model reproduces or aligns with these parameters. The Arrhenius parameters for bulk and various GB types employed in the subsequent analysis are summarized in \cref{Tab:Values}. The values corresponding to bulk, $\Sigma 5$, and HAGBs are adopted from existing literature sources \cite{Jiang2021,Aagesen2022}. For LAGBs, the parameters are assumed to be identical to those of the bulk, as a simplifying approximation. Similarly, the values for other CSL GBs are approximated by those of the $\Sigma 5$ boundary, while the parameters for the $\Sigma 27$ GB are assumed to be equivalent to those of the $\Sigma 9$ GB. For GB types, which are assumed to have the same diffusivity parameters, their volume fractions are added. The new volume fractions are also shown in \cref{Tab:Values}. The Arrhenius plots for all considered GB types and the bulk are shown in \cref{Fig:ArrheniusAll}, which reveals that the $\Sigma 9$ GBs exhibit the highest diffusivity across the temperature range, followed by the $\Sigma 5$ GBs.

\begin{table}[h!]
\centering
\footnotesize
\caption{Arrhenius parameters for Ag diffusion in bulk 3C‑SiC and the GB types considered in this study. For the $\Sigma9$ GB, the parameters associated with the C-rich variant are adopted to represent its behavior, as it constitutes a conservative estimate. As outlined earlier, $f_\text{GB} = 0.0029$.}
\label{tab:ArrheniusParams}
\begin{tabular}{lccc}
\hline
 & $D_0$ [m$^{2}$/s] & $Q$ [eV] & Volume fraction \\
\hline
Bulk 3C-SiC \cite{Jiang2021} & $2.4 \times 10^{-4}$ & 5.34 & $1 - f_\text{GB}$ \\
$\Sigma3$ GBs (this work) & $3.69 \times 10^{-9}$ & 1.02 & $0.46 f_\text{GB}$ \\
HAGBs \cite{Jiang2021} & $1.813 \times 10^{-7}$ & 2.178 & $0.30 f_\text{GB}$ \\
LAGBs (same as the bulk) & $2.4 \times 10^{-4}$ & 5.34 & $0.11 f_\text{GB}$ \\
$\Sigma5$/Other CSL GBs \cite{Aagesen2022} & $2.567 \times 10^{-9}$ & 0.847 & $0.07 f_\text{GB}$ \\
$\Sigma9/\Sigma27$ GBs (this work) & $4.50 \times 10^{-8}$ & 0.88 & $0.06 f_\text{GB}$ \\
\hline
\end{tabular}
\label{Tab:Values}
\end{table}

Applying the effective diffusivity model defined in \cref{Eq:Dav}, with input parameters listed in \cref{Tab:Values}, yields an apparent activation energy of $Q_{\text{eff}} = 0.90$~eV and a pre-exponential factor of $D_{0,\text{eff}} = 6.05 \times 10^{-12}$~m$^2$/s. These values fall outside the 95\% credible intervals of $D_{0,\text{eff}}$ and $Q_\text{eff}$ estimated from the Bayesian analysis (see \cref{Tab:Posterior}), and lead to overprediction of Ag diffusivity. This discrepancy likely stems from missing microstructural effects, notably nano-porosity introduced during CVD processing of SiC \cite{Lopez-Honorato2010,Lopez-Honorato2011,Xiao2012}. The baseline model assumes idealized GB diffusion and neglects mechanisms such as reversible trapping at isolated nano-pores. To capture this missing effect, we incorporate a phenomenological correction factor, $\alpha$, to the trap-free diffusivity:
\begin{equation}
D_\text{eff}' = \alpha D_\text{eff} = \alpha_0 \exp\left( -\frac{Q_t}{kT} \right) D_\text{eff}.
\end{equation}
This form is physically motivated by Shewmon’s derivation \cite{Shewmon2016} showing that the effect of trapping can be fully represented by a multiplicative factor applied to the trap-free diffusivity. Experimental and computational evidence support the existence of such reversible traps in CVD-SiC. López-Honorato \textit{et al.} \cite{Lopez-Honorato2010,Lopez-Honorato2011} reported that Ag atoms in TRISO coatings occupy both GBs and isolated nano‑pores. Xiao \textit{et al.}~\cite{Xiao2012} found that Ag strongly adsorbs onto SiC (001) surfaces and diffuses along them with barriers $\leq 0.5$ eV, but desorption requires, on average, 1.27 eV. Since surface diffusion is fast, long-range transport involving pores is bottlenecked by desorption. Assigning $Q_t$ = 1.27~eV and applying the correction increases the effective activation energy to $Q_\text{eff}'$ = 2.17~eV, which falls within the 95\% credible interval and is close to the posterior peak at 2.30~eV.

To estimate the prefactor, $\alpha_0$, consider a system in which Ag atoms are initially immobilized in traps (e.g., nanopores) within a GB network and must overcome two successive barriers to contribute to long-range diffusion. The first is a desorption barrier, $Q_t$, which allows the atom to leave the trap, and the second is a \textit{commitment barrier}, $Q_{\text{eff}}$, which determines whether the atom enters the diffusive GB network or returns to the trap. The model is based on the transition sequence:
\begin{equation}
T \xrightarrow{k_1} F \xrightarrow{k_2} M, \quad \text{with} \quad F \xrightarrow{k_{-1}} T,
\end{equation}
where $T$ is the trap state, $F$ is a short-lived intermediate free state near the pore suface \cite{Xiao2012}, and $M$ is the mobile state within the GB. The short-lived free state, $F$, has been formalized by Xiao \textit{et al.}'s DFT study of Ag surface diffusion in 3C-SiC.

The desorption rate from state $T$ is given by transition-state theory as:
\begin{equation}
k_1 = \nu_0 \exp\left( -\frac{Q_t}{kT} \right),
\end{equation}
where $\nu_0 \sim 10^{13}$ s$^{-1}$ is the attempt frequency \cite{Sholl2009}. Similarly, the commitment rate is:
\begin{equation}
k_2 = \nu_0 \exp\left( -\frac{Q_{\text{eff}}}{kT} \right),
\end{equation}
and we assume $k_{-1} \approx \nu_0$, based on the approximation that fall-back to the trap is barrier-less.

After desorption, the atom enters the intermediate state $F$, from which it may either return to the trap with rate $k_{-1}$ or commit to the mobile state with rate $k_2$. The total rate of leaving state $F$ is $k_{-1} + k_2$. The probability that the next transition is $F \to M$ rather than $F \to T$ is then the conditional probability that the commitment transition occurs before the fall-back transition:
\begin{equation}
P_{\text{commit}} = \frac{k_2}{k_{-1} + k_2}.
\end{equation}
Substituting the expressions for the rates gives:
\begin{equation}
P_{\text{commit}} = \frac{ \nu_0 \exp(-Q_{\text{eff}} / (kT))}{ \nu_0 + \nu_0 \exp(-Q_{\text{eff}} / (kT))} = \frac{1}{1 + \exp(Q_{\text{eff}} / (kT))}.
\end{equation}
In the limit $Q_{\text{eff}} \gg kT$, which is typical for strongly activated processes at low to moderate temperatures, the exponential term dominates the denominator, and we may approximate:
\begin{equation}
P_{\text{commit}} \approx \exp\left( -\frac{Q_{\text{eff}}}{kT} \right).
\end{equation}

A diffusion cycle is the sequence in which an Ag atom desorbs from a nanopore, migrates along the GB network, and is eventually retrapped at another pore. Once an atom commits to the GB network, it performs a series of unbiased hops of length $\lambda$ until it is captured by another trap. The MSD during such a successful episode is $\langle h \rangle \lambda^2$. However, not all desorption attempts result in commitment. The probability of success is $P_{\text{commit}}$, so the average mean squared displacement per desorption attempt is:
\begin{equation}
\langle \Delta r^2 \rangle = P_{\text{commit}} \cdot \langle h \rangle \lambda^2.
\end{equation}

Assume the total time per diffusion cycle is dominated by the residence time in the trap:
\begin{equation}
t_{\text{cycle}} \approx \tau_{\text{trap}} \approx \frac{1}{k_1} = \tau_0 \exp\left( \frac{Q_t}{kT} \right),
\end{equation}
with $\tau_0 = 1/\nu_0$. During a successful escape, the atom undergoes a random walk with an average number of hops $\langle h \rangle$ before retrapping. If the probability of encountering a trap per hop is $P_\text{trap}$, then the number of hops before retrapping is a geometric random variable \cite{Steyer2017} with mean:
\begin{equation}
\langle h \rangle = \frac{1}{P_\text{trap}}.
\end{equation}

The effective diffusivity is obtained by combining the MSD per cycle with the time required to complete one full cycle. Using the Einstein relation for diffusivity in $d$ dimensions, the effective trap-limited diffusivity becomes:
\begin{equation}
D_{\text{eff}}' = \frac{\langle \Delta r^2 \rangle}{2d \, \tau_{\text{trap}}} = \frac{P_{\text{commit}} \langle h \rangle \lambda^2}{2d \, \tau_0} \exp\left( -\frac{Q_t}{kT} \right).
\end{equation}
Substituting the approximation $P_{\text{commit}} \approx \exp(-Q_{\text{eff}} / (kT))$ yields:
\begin{equation}
D_{\text{eff}}' = \frac{\langle h \rangle \lambda^2}{2d \, \tau_0} \exp\left( -\frac{Q_t + Q_{\text{eff}}}{kT} \right).
\end{equation}
Defining the trap-free diffusivity prefactor as:
\begin{equation}
D_{0,\text{eff}} = \frac{\lambda^2}{2d \, \tau_0}.
\end{equation}
The final expression for the effective trap-limited diffusivity becomes:
\begin{equation}
D_{\text{eff}}'(T) = \langle h \rangle D_{0,\text{eff}} \exp\left( -\frac{Q_t + Q_{\text{eff}}}{kT} \right).
\end{equation}

It now remains to estimate $\langle h \rangle$. Assuming spherical grains of radius $R_g = 500$~nm, pores of radius $R_p$, and porosity $\phi$, the pore number density is:
\begin{equation}
n_p = \frac{\phi}{\tfrac{4}{3} \pi R_p^3},
\end{equation}
and the grain number density is:
\begin{equation}
n_g = \frac{1}{\tfrac{4}{3} \pi R_g^3}.
\end{equation}
The number of pores per grain is then:
\begin{equation}
N = \frac{n_p}{n_g} = \phi \left( \frac{R_g}{R_p} \right)^3.
\end{equation}

We estimate $P_\text{trap}$ by approximating the pore-affected GB length per grain as $N \cdot 2 R_p$ and the total GB perimeter per grain as $2\pi R_g$:
\begin{equation}
P_\text{trap} = \frac{N \cdot 2R_p}{2\pi R_g} = \frac{N R_p}{\pi R_g}.
\end{equation}
Then,
\begin{equation}
\langle h \rangle = \frac{\pi R_g}{N R_p} = \frac{\pi}{\phi} \left( \frac{R_p}{R_g} \right)^2.
\end{equation}
This gives the final equation for the trap-limited effective diffusivity of silver in 3C-SiC as:
\begin{equation}
D_\text{eff}' = \frac{\pi}{\phi} \left( \frac{d_p}{d_g} \right)^2 D_{0,\text{eff}} \exp\left(-\frac{Q_\text{eff}+Q_t}{kT} \right),
\end{equation}
where $d_p$ and $d_g$ are the average pore and grain sizes, respectively, $\phi$ is the porosity, $Q_t$ is the desorption energy, and $D_{0,\text{eff}}$ and $Q_\text{eff}$ are the parameters of the effective trap-free diffusivity (\cref{Eq:Dav}).

SiC for TRISO particles is produced with low porosity within the range $\phi$ = 0.25--1.50\% \cite{Rohbeck2016}, and pore radii spanning $R_p$ = 50--350 nm \cite{Bari2013,Lopez-Honorato2010,Lopez-Honorato2011}. Using mid-range representative values ($\phi = 0.875\%$, $R_p = 200$~nm, and $R_g = 500$~nm), we find $\langle h \rangle \approx 57$. Note that across the ranges of $\phi$ and $R_p$, $\langle h \rangle$, varies over two orders of magnitude, from $\mathcal{O}(1)$ to $\mathcal{O}(10^2)$, with the representative value of 57 corresponding to the average order of magnitude over this range. The product $\alpha_0 D_{0,\text{eff}}$ = $3.45 \times 10^{-10}$~m$^2$/s closely matches the mode value of $3.03 \times 10^{-10}$~m$^2$/s obtained from the Bayesian posterior.

Although approximate and based on qualitative arguments, the proposed correction introduces a physically motivated mechanism that reconciles the model predictions with experimental diffusivities. It suggests that GB diffusion and pore-mediated migration act in concert, not in competition, in typical CVD-SiC microstructures. The formulation holds provided pores are sparse and isolated, so that vapor-phase or surface-mediated transport does not dominate. At higher porosity or pore connectivity, alternative models incorporating interconnected surface diffusion or gas-phase mechanisms would be necessary.

\subsection{Local sensitivity analysis}

\begin{figure}[h!]
 \centering
 \includegraphics[width=0.7\textwidth]{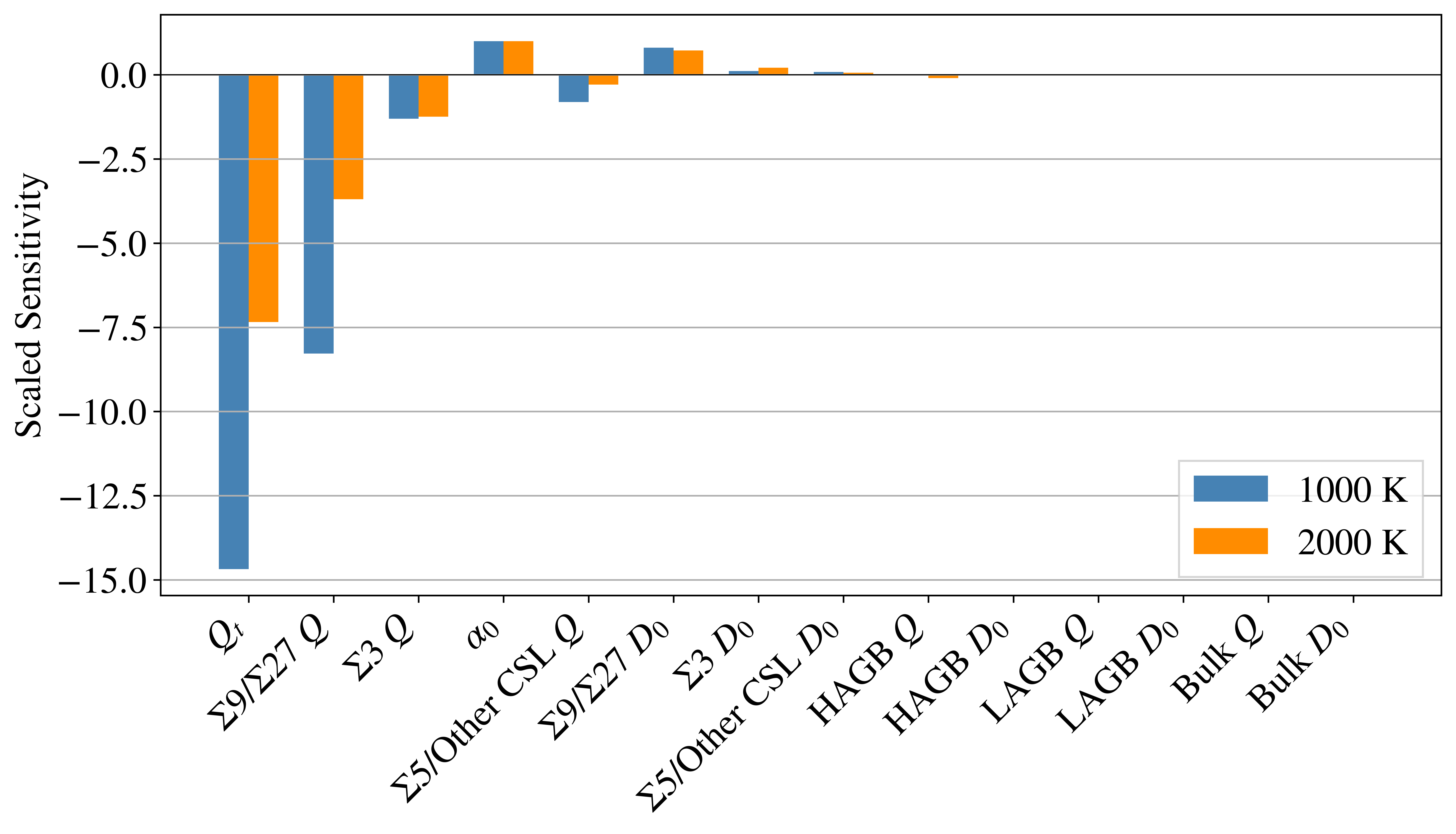}
 \caption{Scaled local sensitivity indices of $D_ \text{eff}$ at 1000 K and 2000 K. These indices reflect the effect of infinitesimal perturbations around nominal parameter values.}
 \label{Fig:Sensitivity}
\end{figure}

To assess the parametric control over the diffusivity model, we evaluate the scaled local sensitivities of $D_{\mathrm{eff}}$ with respect to all underlying Arrhenius parameters $D_{0,i}$, $Q_i$, as well as the multiplicative model parameters $\alpha_0$ and $Q_t$, at 1000~K and 2000~K. The sensitivity indices are shown in \cref{Fig:Sensitivity}.

Among all parameters, $Q_t$ exhibited the strongest influence, with scaled sensitivities of approximately $-14.68$ at 1000~K and $-7.34$ at 2000~K. This large negative sensitivity is a direct consequence of the exponential prefactor $\exp(-Q_t / (k T))$ that uniformly modulates the entire diffusivity expression. Mathematically, the sensitivity is governed by the derivative $\partial D_{\mathrm{eff}} / \partial Q_t \sim -1/(k T)$, and thus scales inversely with temperature. This explains why the sensitivity of $Q_t$ diminishes by nearly a factor of two at 2000~K relative to 1000~K. The physical implication is that even modest uncertainties in $Q_t$ can induce order-of-magnitude deviations in $D_\text{eff}$, especially in the lower-temperature regime. The scaling parameter $\alpha_0$, in contrast, exhibited a constant sensitivity of exactly $1$ at both temperatures. This is consistent with its role as a simple multiplicative factor in the model.

Within the GB structure, the $\Sigma 9/\Sigma 27$ GB demonstrates the most prominent effect among the microstructural components. Its activation energy $Q$ had scaled sensitivities of $-8.28$ at 1000~K and $-3.70$ at 2000~K, while its pre-exponential factor $D_0$ had corresponding sensitivities of $+0.81$ and $+0.72$. Notably, these values are larger than those of $\Sigma 3$, despite the latter contributing a significantly higher volume fraction ($\sim 46\%$ vs.\ $\sim 6\%$). This apparent discrepancy arises from the much higher intrinsic diffusivity of $\Sigma 9/\Sigma 27$ GBs due to their large $D_0$ and low $Q$, causing them to dominate the harmonic mean that represents the lower bound of diffusivity, $D_l$.

The $\Sigma 3$ GB, while more abundant, showed moderate sensitivity: $-1.30$ and $-1.25$ for $Q$, and $+0.11$ and $+0.21$ for $D_0$ at 1000~K and 2000~K, respectively. This reflects a dual effect: $\Sigma 3$ is relatively fast but not fast enough to match $\Sigma 9/\Sigma 27$, and its impact becomes more visible at elevated temperatures where the differences between GB diffusivities narrow. This trend reinforces the understanding that in harmonic averaging, the fastest components, regardless of fraction, tend to control the overall effective property.

Other GB types, such as $\Sigma 5$/Other CSL and HAGBs, show limited impact. This stems from their intermediate diffusivities and modest volume fractions, which prevent them from becoming dominant contributors to the effective transport. The bulk and LAGBs show virtually zero sensitivity across the board, which is expected. In a harmonic mean, such slow pathways act as bottlenecks but do not define the transport rate unless they represent a majority fraction, which is not the case here.

Overall, the temperature-dependent behavior of the sensitivities highlights that exponential terms dominate at lower temperatures. At higher temperatures, differences in $Q$ become less pronounced, and the pre-exponential factors gain slightly more influence. The model structure amplifies these effects, especially through the presence of a global prefactor involving $Q_t$, which governs the entire curve shape and temperature scaling. These results suggest that controlling the population of high-diffusivity $\Sigma 9/\Sigma 27$ GBs could meaningfully limit Ag transport. Also, accurate characterization of global scaling terms like $Q_t$ is critical for reliable model prediction. The sensitivity framework thus provides a rigorous basis for prioritizing parameter calibration and guiding microstructural engineering.


\section{Discussion}

This study presents a multi-paradigm modeling framework for predicting Ag transport in polycrystalline 3C–SiC. The central contribution lies in integrating MD simulations, bounds averaging, and uncertainty quantification to derive a mechanistically interpretable expression for the effective diffusivity of silver that incorporates the distinct contributions of specific GB types and accounts for dilute trapping at nano-porosity. Specifically, we compute Ag diffusivities in $\Sigma 3$ and $\Sigma 9$ GBs using MD, supplementing these with literature-derived values for other GB types. These diffusivities are combined using a physically motivated bounds-averaging scheme that captures the influence of both transport anisotropy and microstructural heterogeneity.

Several limitations and simplifying assumptions merit discussion. First, this work considers only GB-mediated transport and dilute nano-porosity. Other microstructural features, irradiation effects, or chemical interactions are not treated. In particular, the model neglects Ag–Pd interactions, which have been suggested to play a role in silver mobility under certain experimental conditions. Second, porosity is considered only in the dilute limit. We assume that nano-pores act as isolated, reversible traps and do not contribute directly to long-range percolative transport. Consequently, the proposed framework breaks down at high porosity concentrations or under conditions where pore connectivity is significant. In such regimes, surface- or vapor-mediated transport would likely dominate, violating the central assumption of GB-governed diffusion. However, the dilute-porosity assumption aligns with experimental characterizations of CVD-fabricated SiC, where nano-pores are non-interconnected and separated by at least one grain diameter.

Our model also introduces a physically justified multiplicative correction factor to account for desorption-limited transport at isolated pore surfaces. We derive this term mechanistically by linking the desorption barrier to the number of hops an atom performs between successive trapping events. This correction reconciles the bounds-averaged diffusivity with experimental measurements.

Simon \textit{et al.}~\cite{Simon2022} proposed an alternative mesoscale diffusion model that blends atomistic input with continuum-scale resolution. However, their model treats all GBs as generic HAGBs, omitting the critical influence of specific CSL types such as $\Sigma 3$ and $\Sigma 9$. Moreover, their correction for artificial GB widening is heuristic in nature and lacks a rigorous physical basis. In contrast, our model systematically incorporates GB character, volume fraction, and diffusivity, and is calibrated against experimental observations using a physically motivated and mechanistically interpretable trapping correction. This results in a more granular and physically faithful model of Ag transport in 3C–SiC.

At the engineering scale, Dhulipala \textit{et al.} \cite{Dhulipala2024,Dhulipala2025} have shown that model inadequacy dominates uncertainty in Ag release simulations for TRISO particles, even when lower length-scale (LLS) inputs are used. Notably, they conclude that improved physical fidelity at the mesoscale is needed to reduce predictive error. The present work directly addresses this need by refining the treatment of Ag diffusivity in 3C-SiC and introducing an explicit model for the interaction between GBs and isolated nano-porosity. Our formulation thus targets one of the key contributors to system-level model inadequacy identified in their studies.

A natural next step would be to implement our derived effective diffusivity model into a fuel performance code such as BISON \cite{Williamson2021}. Conducting a follow-up uncertainty quantification, along the lines of Dhulipala \textit{et al.}, would then allow for a quantitative evaluation of whether this refined, physics-based model can meaningfully reduce the overall model inadequacy in engineering-scale simulations of Ag release.

\section{Conclusions}

We have developed a physics-informed model for estimating the effective diffusivity of silver in polycrystalline 3C–SiC by integrating MD simulations, analytical homogenization, and Bayesian calibration. Diffusivities in $\Sigma 3$ and $\Sigma 9$ GBs were obtained from MD, while values for other GB types and bulk were drawn from the literature. These were combined using a bounds-averaging approach that accommodates multiple GB types with distinct transport characteristics. A Bayesian analysis of experimental Ag diffusivity data was performed to infer credible intervals for the effective Arrhenius parameters and to establish a physically motivated correlation between activation energy and pre-exponential factor.

While the homogenized model provides a mechanistic description of GB-mediated transport, it systematically overestimates silver diffusivity relative to experimental data. To reconcile this discrepancy, we introduced a multiplicative correction based on reversible trapping at isolated nano-pores. The form and magnitude of this correction were derived from first-principles arguments and shown to yield effective transport parameters consistent with experimental observations. This treatment naturally captures the experimentally observed coexistence of GB and surface-mediated transport.

Sensitivity analysis revealed that Ag transport is most strongly influenced by the activation energy associated with trap desorption and by the diffusivity of $\Sigma 9$ GBs, which, despite their limited volume fraction, dominate the transport due to their high intrinsic diffusivity. These results highlight the importance of accurately characterizing GB populations and of accounting for the presence of dilute nano-porosity in predictive models of fission product transport.

The framework developed in this work enables a mechanistically grounded, closed-form description of Ag diffusivity in SiC that can be directly embedded in higher-scale fuel performance simulations. Future work will focus on implementing this model into finite element codes and assessing whether the improved mesoscale fidelity translates into reduced model inadequacy in TRISO fuel behavior predictions.

\section{Acknowledgments}

The authors acknowledge the support of EPRI on this project through their participation in the Consortium for Nuclear Power at North Carolina State University. This research made use of the resources of the High-Performance Computing Center at Idaho National Laboratory, which is supported by the Office of Nuclear Energy of the U.S. Department of Energy and the Nuclear Science User Facilities under Contract No. DE-AC07-05ID14517. Mohamed AbdulHameed is grateful to Mostafa Hamza for the useful discussions and suggestions. This work is dedicated to Mohamed Emara, the physics teacher who introduced Mohamed AbdulHameed to electric circuits---knowledge that, more than a decade later, proved useful in modeling impurity transport.

\appendix

\section{Derivation of $D_{\Sigma 5}$}
\label{app}

The mean squared displacement in two dimensions can be expressed as:
\begin{equation}
\langle (\Delta r)^2 \rangle = \langle (\Delta x)^2 \rangle + \langle (\Delta y)^2 \rangle.
\end{equation}
Assuming Brownian motion along each axis with diffusivities $D_x$ and $D_y$, the displacements satisfy:
\begin{equation}
\langle (\Delta x)^2 \rangle = 2 D_x t, \quad \langle (\Delta y)^2 \rangle = 2 D_y t.
\end{equation}
Substituting into the expression for $\langle (\Delta r)^2 \rangle$ gives:
\begin{equation}
\langle (\Delta r)^2 \rangle = 2(D_x + D_y)t.
\end{equation}
Alternatively, one may define an effective two-dimensional diffusivity $D_{xy}$ via:
\begin{equation}
\langle (\Delta r)^2 \rangle = 4 D_{xy} t,
\end{equation}
which yields:
\begin{equation}
D_{xy} = \frac{D_x + D_y}{2}.
\label{Eq:Dxy}
\end{equation}

This derivation relies on several assumptions. First, displacements along orthogonal directions are assumed to be statistically independent, so that cross-terms like $\langle \Delta x \Delta y \rangle$ vanish. Second, motion along each axis is modeled as Brownian. These assumptions, combined with the linearity of the expectation operator, justify expressing the total MSD as a sum of individual components. This formulation is valid for both isotropic and anisotropic systems as long as the above conditions hold.

\bibliographystyle{elsarticle-num}
\bibliography{ref}

\end{document}